\documentclass[11pt]{article}

\usepackage{natbib}

\usepackage{amsmath}
\usepackage{amssymb}
\usepackage{amsthm}
\usepackage{bbm}
\usepackage{bm}

\usepackage{graphicx}

\usepackage{apalike}
\usepackage{booktabs}
\usepackage{caption}
\usepackage{subcaption}
\usepackage{a4wide}
\usepackage{multirow}
\usepackage{booktabs}
\usepackage{siunitx}
\usepackage{color,soul}

\usepackage{pdflscape}
\usepackage{afterpage}
\usepackage{capt-of}

\usepackage{lipsum}

\usepackage{changes}
\usepackage{xcolor}
\newcommand{\stkout}[1]{\ifmmode\text{\sout{\ensuremath{#1}}}\else\sout{#1}\fi}
\setdeletedmarkup{\stkout{#1}}

\theoremstyle{definition}
\newtheorem{theorem}{Theorem}
\newtheorem{lemma}[theorem]{Lemma}

\newtheorem{proposition}[theorem]{Proposition}
\newtheorem{remark}[theorem]{Remark}

\newtheorem{example}{Example}

\newcommand{\E}{\mathbb E}
\newcommand{\e}{\mathrm e}

\newcommand{\D}{\mathrm{d}}

\newcommand{\vc}{\mathrm{vec}}

\newcommand{\PP}{\mathbb P}
\newcommand{\Var}{\mathrm{Var}}

\newcommand{\Dg}{\mathrm{Dg}}

\newcommand{\A}{\bm{\mathrm{A}}}
\newcommand{\B}{\bm{\mathrm{B}}}

\begin{document}
	
\title{Multi-kernel property in high-frequency price dynamics under Hawkes model}

\author{Kyungsub Lee\footnote{Department of Statistics, Yeungnam University, Gyeongsan, Gyeongbuk 38541, Republic of Korea, Corresponding author, Email: ksublee@yu.ac.kr}}

\date{}

\maketitle	

\begin{abstract}
This study investigates and uses multi-kernel Hawkes models to describe a high-frequency mid-price process.
Each kernel represents a different responsive speed of market participants.
Using the conditional Hessian, we examine whether the numerical optimizer effectively finds the global maximum of the log-likelihood function under complicated modeling.
Empirical studies that use stock prices in the US equity market show the existence of multi-kernels classified as ultra-high-frequency (UHF), very-high-frequency (VHF), and high-frequency (HF).
We estimate the conditional expectations of arrival times and the degree of contribution to the high-frequency activities for each kernel.
\end{abstract}


\section{Introduction}

In recent years, interest in financial activities under the high-frequency level has increased as time records for financial market transactions, 
and quotes are achieved at higher time resolutions.
Intraday events in major stock markets were recorded
in milliseconds previously, and then the resolution gradually increases to microseconds and nanoseconds. 
High-frequency financial data comprise the actions of ultra-high-frequency and high-frequency traders as well as low-frequency participants in financial markets.

The Hawkes process \citep{Hawkes1}, which has been applied to the natural and social sciences, 
captures the jump increment, called excitement, in intensities originating from past events and decays of the effect.
Activities in financial markets under a tick structure, 
where prices vary tick-by-tick, 
such as transactions, quote revisions, and cancellations
are activated by past events; 
their aftermath effects diminish over time.
Due to this commonality, 
the Hawkes model has been actively used as a financial model in recent years.
For the earliest studies,
\cite{Bowsher2007} proposed a generalized Hawkes model for multivariate financial market events, 
which incorporates various dependencies in market features,
 and \cite{Large2007} utilized a mutually exciting Hawkes model to describe the resiliency in the limit order book.

Although the Hawkes model can be used in various fields of finance, such as credit risk analysis \citep{Errais2010,AITSAHALIA2015585,ma2016structural,ketelbuters2022cds} and optimal execution \citep{choi2021,da2021simple,jusselin2021optimal,gavsperov2022deep},
studies focusing on price dynamics, such as our study, are also abundant.
To mention a few, \cite{Bacry2013modelling} used the multivariate Hawkes model for the tick dynamics of asset prices with the applications of microstructure noise analysis.
Under settings similar to the standard Hawkes models, studies by
\cite{Bacry2014,DaFonseca&Zaatour2014,da2017correlation,arouri2019cojumps,ji2020combining} 
contributed to the literature in aspects such as stylized facts of market impact, moment analysis, lead-lag relationship, capturing correlated jumps, extreme risk, and so on.

In addition, Hawkes models of a modified and extended form suitable for more complex structures have also been proposed.
For example, to describe the bid and ask price dynamics, 
\cite{Zheng2014} introduced a multivariate constraint Hawkes-like point process
and \cite{LeeSeo2022} proposed an extended Hawkes model to capture the spread dependent intensities.
\cite{Jang2020} suggested the Hawkes flocking model to assess the systemic risk in crude oil and gasoline futures markets.
\cite{morariu2021state} examined state-dependent Hawkes processes with an application to limit order book dynamics.
\cite{swishchuk2021modelling} proposed general regime-switching  Hawkes models to describe the price processes in the limit order books.
\cite{Zhang2021} suggested a new model by combining a Hawkes process and a finite-range contact process for stock price movements.

For reviews of other interesting studies on the Hawkes model with financial applications 
that are been mentioned here, please refer to \cite{Law2015}, \cite{Bacry2015hawkes}, and \cite{Hawkes2018}.

We focus on the multi-kernel property in the high-frequency mid-price process in stock and futures markets.
The mid-price, the mean of the best bid and ask prices, 
moves in multiples of the minimum tick in market microstructure.
As mentioned before, many studies have modeled price movement in the tick structure using the Hawkes model.
Among these, a single kernel is mainly used, as in \cite{Bacryetal2013,Fonseca2014,lee2017modeling}
or power-law kernel, as in \cite{hardiman2013critical,filimonov2015apparent,bacry2016estimation}.
However, relatively few studies have been conducted on multi-kernel exponential models.

The intraday movements of the stock price are significantly influenced by ultra-high-frequency traders.
However, there are also numerous activities that are performed in relatively lower frequencies.
The responsiveness to previous events differs depending on their computing power, network speed, and available services such as colocation.
The multi-kernel Hawkes model can capture the different responsive speeds and categorize the different types of activities in financial activities.
We examine the differences in responsiveness among traders verified by multiple kernels.

As a basic property,
we derive the first and second moments under the bi-variate multi-kernel Hawkes model,
which is an extension of the results in a single kernel model, such as in \cite{DaFonseca&Zaatour2014}, \cite{lee2017modeling}, and \cite{cui_hawkes_yi_2020}.
The moment expression is a closed-form formula
and it helps verify the accuracy of the simulation of the multi-kernel Hawkes model.

We carefully examine the estimation procedure 
because the number of parameters increases as the number of kernels increases.
We discuss the property of the global maximum of the log-likelihood based on a conditional Hessian to examine the reliability of the multivariate and multi-kernel model estimation.
Various simulation examples are also presented.


The remainder of this paper is organized as follows.
Section~\ref{sect:model} proposes the multi-kernel Hawkes model for mid-price processes and discusses the global maximum property and moments.
Section~\ref{sect:result} present an empirical study based on high-frequency stock prices and the filtering effect.
Finally, Section~\ref{sect:concl} concludes the paper.

\section{Model}\label{sect:model}

\subsection{Multi-kernel model in high-frequency data}

The mid-price processes of financial assets exhibit typical signs of high-frequency trading.
The price reacts rapidly to previous events and the impact dissipates rapidly.
However, market participants may have different processing speeds.
We separate the kernels based on the frequency spectrum by introducing multi-kernels in the Hawkes model.
It has often been argued that ultra-high-frequency financial raw data recorded at a high-resolution are not adequately suited to the single-kernel exponential Hawkes model.
Note that stock movements have been recorded in nanoseconds in major stock exchanges since approximately 2015.
Given that there are few in-depth studies on the financial application of the multi-kernel exponential Hawkes model,
through this study, we intend to identify the advantages of the multi-kernel model as well as its disadvantages.

Consider a bi-variate counting process and its intensities:
$$ 
\bm{N}_t =
\begin{bmatrix} 
N_1(t) \\
N_2(t) \\
\end{bmatrix}, \quad
\bm{\lambda}_t = 
\begin{bmatrix} 
\lambda_1(t) \\
\lambda_2(t) \\
\end{bmatrix}
$$
where $N_1(t)$ and $N_2(t)$ count the number of the up and down movements of a mid-price process up to $t$, respectively.
Thus, the mid-price process can be represented as the difference between the two counting processes.
In this study, bold face letters indicate vectors or matrices, as we deal with the bi-variate model.
Let the vector of the intensity process be
\begin{equation}
\bm{\lambda}_t = \bm{\mu} + \int_{-\infty}^{t} \bm{h}(t-u) \D \bm{N}_u\label{Eq:lambda}
\end{equation}
where $\bm{\mu} = [\mu_1, \mu_2]^{\top}$ is constant.

This study focuses on the multi-kernel property,
which is different from other studies on the single kernel model; that is
$$ \bm{h} = \sum_{k=1}^{K} \bm{h}_k $$
where $k$ denotes the kernel number.
Each kernel is
$$ 
\bm{h}_k (t) = \bm{\alpha}_k \circ \begin{bmatrix}
\e^{-\beta_{k11} t} & \e^{-\beta_{k12} t} \\
\e^{-\beta_{k21} t} & \e^{-\beta_{k22} t}
\end{bmatrix}.
$$
where $\bm{\alpha}_k$ is a $2\times 2$ matrix whose element in $i$-th row and $j$-th column are represented by $\alpha_{ijk}.$
The smaller the $k$, the higher the kernel frequency,
hence the larger the $\beta_k$.

As the number of parameters increases proportionally with the number of kernels, 
for the sake of model parsimony, 
we can assume that $\beta$s are the same for each kernel or for each raw of a kernel depending on the context.
Multiple kernels are designed to capture different responsiveness speeds for high-frequency activities in the financial market.

\subsection{Moment property}~\label{Subsect:moment}

In this section, we computes the moments using the bi-variate multi-kernel Hawkes model.
The intensity process of the multi-kernel Hawkes process can be rewritten as
\begin{equation}
	\bm{\lambda}_t = \bm{\mu} + \sum_{k=1}^{K}\bm{\lambda}_{k}(t) \label{Eq:multi}
\end{equation}
where the components are defined by
$$ \bm{\lambda}_{k}(t) = \begin{bmatrix}
	\lambda_{k1}(t) \\
	\lambda_{k2}(t)
\end{bmatrix} =
\int_{-\infty}^{t}\bm{h}_k(t-u) \D \bm{N}_u.
$$
To derive these formulas, we have additional restrictions on the Markov property:
$$ \beta_{k1} := \beta_{k11} = \beta_{k12}, \quad \beta_{k2} := \beta_{k21} = \beta_{k22},$$
i.e.,
$$
\bm{h}_k(t) = \bm{\alpha}_k \circ
\begin{bmatrix}
	\e^{-\beta_{k1} t} & \e^{-\beta_{k1} t} \\
	\e^{-\beta_{k2} t} & \e^{-\beta_{k2} t}
\end{bmatrix}.
$$
Following this, the differential form of the intensity process can be represented by
\begin{align}
	&\D \bm{\lambda}_t = \sum_{k=1}^{K} \D \bm{\lambda}_{k}(t), \\
	&\D \bm{\lambda}_{k}(t) = - \bm{\beta}_k \bm{\lambda}_{k}(t) \D t + \bm{\alpha}_k \D \bm{N}_t \label{Eq:dlk}
\end{align}
where
$$  \bm{\beta}_k  = \begin{bmatrix} \beta_{k1} & 0 \\ 0 & \beta_{k2}\end{bmatrix}.$$
For the simplicity of the formula, we assume that the processes are in a steady state at time 0, 
that is, by considering the intensity process to begin at $-\infty$, 
we always assume that the unconditional distribution of the intensity processes at time 0 are in a steady state.
Under this assumption, 
the expectation of the intensity processes and their components are constant.

\begin{proposition}~\label{Prop:1}
	Under the steady state assumption, for $t>0$,
	\begin{align*}
		&\E[\bm{\lambda}_{t}] = \left(\bm{\mathrm{I}} - \sum_{k=1}^{K}\bm{\beta}_k^{-1} \bm{\alpha}_k \right)^{-1} \bm{\mu} ,\\
		&\E[\bm{\lambda}_{k}(t)] = \bm{\beta}_k^{-1} \bm{\alpha}_k \E[\bm{\lambda}_t].
	\end{align*}
\end{proposition}

To proceed further, define
$$ \bm \Lambda_t = \begin{bmatrix} \bm{\lambda}_{1}(t) \\ \vdots \\ \bm{\lambda}_{K}(t) \end{bmatrix}, \quad  \bm{\alpha} = \begin{bmatrix} \bm{\alpha}_1 \\ \vdots \\ \bm{\alpha}_K \end{bmatrix}, \quad \bm{\beta} = 
\Dg([\beta_{11}~ \beta_{12}~\cdots ~\beta_{K1}~ \beta_{K2}])$$
where $\mathrm{Dg}(\cdot)$ denotes a diagonal matrix whose diagonal entries are elements in the argument.
Thus, $\Lambda_t$ is a $2K\times 1$ matrix, and $\bm{\alpha}$ is $2K\times 2$ and $\bm{\beta}$ is a diagonal $2K \times 2K$ matrix.
Following this,
$$ \D \bm \Lambda_t = \bm{\beta} \bm \Lambda_t \D t + \bm{\alpha} \D \bm{N}_t.$$

\begin{lemma}\label{Lemma:quad}
	Consider $2\times 1$ vector processes $\bm{X}$ and $\bm{Y}$ such that
	\begin{align*}
		\D \bm{X}_t = \bm{a}_t \D t + \bm{f}_x(t) \D \bm{N}_t, \quad \D \bm{Y}_t = \bm{b}_t \D t + \bm{f}_y(t) \D \bm{N}_t.
	\end{align*}
	Subsequently, the differential form of the quadratic variation matrix process of $\bm{X}\bm{Y}^{\top}$ is represented by
	$$ \D [\bm{X}\bm{Y}^{\top}]_t = \bm{f}_x(t) \mathrm{Dg} (\D \bm{N}_t) \bm{f}_y^{\top}(t).$$
	Thus,
	\begin{align*}
		\D (\bm{X}_t \bm{Y}^{\top}_t) &= \bm{X}_{t-} \D \bm{Y}^{\top}_t + \D(\bm{X}_t) \bm{Y}^{\top}_{t-} + \bm{f}_x(t) \mathrm{Dg} (\D \bm{N}_t) \bm{f}_y^{\top}(t).
	\end{align*}
	In addition,
	\begin{align*}
		\frac{\D\E[\bm{X}_t \bm{Y}^{\top}_t]}{\D t} = \E[ \bm{X}_{t-} (\bm{b}_t^{\top} + \bm{\lambda}_t^{\top} \bm{f}_y^{\top}(t))] + \E[(\bm{a}_t + \bm{f}_x(t)  \bm{\lambda}_t ) \bm{Y}^{\top}_{t-} ] 
		+ \bm{f}_x(t) \mathrm{Dg} (\E[\bm{\lambda}_t]) \bm{f}_y^{\top}(t).
	\end{align*}	
\end{lemma}

\begin{proposition}\label{Prop:syl}
	Under the steady state assumption, $\E[  \bm{\Lambda}_t \bm{\Lambda}_{t}^{\top} ]$ satisfies the following Sylvester equation:
	\begin{equation}
		(\bm{\beta} - \bm{\alpha}\bm{\mathrm{J}})\E[  \bm{\Lambda}_t \bm{\Lambda}_{t}^{\top} ]  + \E[\bm{\Lambda}_{t}  \bm{\Lambda}_{t}^{\top}] (\bm{\beta} - \bm{\mathrm{J}}^{\top} \bm{\alpha}^{\top}) = \bm{\alpha}\bm{\mu} \E [ \bm{\Lambda}_{t}^{\top}]  + \E[\bm{\Lambda}_{t} ]\bm{\mu}^{\top}\bm{\alpha}^{\top} +  \bm{\alpha} \mathrm{Dg}(\E[\bm{\lambda}_t])\bm{\alpha}^{\top}\label{Eq:Ell}
	\end{equation}
	where $\bm{\mathrm{J}}$ is a $2 \times 2K$ matrix such that
	$$ \bm{\mathrm{J}} = \begin{bmatrix} 1 & 0 & \cdots & 1 & 0 \\  0 & 1 & \cdots & 0 & 1 \end{bmatrix}.$$
\end{proposition}

\begin{remark}
	Solving Eq.~\eqref{Eq:Ell} is equivalent to solving
	$$ (\bm{\mathrm{I}} \otimes (\bm{\beta} - \bm{\alpha}\bm{\mathrm{J}}) + (\bm{\beta} - \bm{\alpha}\bm{\mathrm{J}}) \otimes \bm{\mathrm{I}} ) \vc \left( \E[  \bm{\Lambda}_t \bm{\Lambda}_{t}^{\top} ] \right) = \vc \left( \bm{\alpha}\bm{\mu} \E [ \bm{\Lambda}_{t}^{\top}]  + \E[\bm{\Lambda}_{t} ]\bm{\mu}^{\top}\bm{\alpha}^{\top} +  \bm{\alpha} \mathrm{Dg}(\E[\bm{\lambda}_t])\bm{\alpha}^{\top} \right)$$
	where $\otimes$ denotes the Kronecker product, and $\vc$ is the vectorization operator.
\end{remark}

\begin{proposition}\label{Prop:lN}
	We have
	$$\E[\bm{\lambda}_t \bm{N}_t^{\top}] \approx \A t + \B $$
	where
	\begin{align}
		\A &= \left(\bm{\mathrm{I}} - \sum_{k=1}^{K} \bm{\beta}_k^{-1}\bm{\alpha}_k\right)^{-1} \bm{\mu} \E[\bm{\lambda}_t^{\top}]  = \E[\bm{\lambda}_t] \E[\bm{\lambda}_t^{\top}]\label{Eq:A}\\
		\B &= \left(\bm{\mathrm{I}} - \sum_{k=1}^{K} \bm{\beta}_k^{-1}\bm{\alpha}_k\right)^{-1} \sum_{k=1}^{K}\bm{\beta}_{k}^{-1} \left( - \bm{\beta}_k^{-1}\bm{\alpha}_k \A + \E[\bm{\lambda}_{k}(t) \bm{\lambda}^{\top}_t] + \bm{\alpha}_k \mathrm{Dg} (\E [\bm{\lambda}_t] ) \right)\label{Eq:B}
	\end{align}
\end{proposition}

\begin{remark}
	Note that
	$$ \bm{\lambda}_{k}(t) \bm{\lambda}^{\top}_t = \bm{\mathrm{J}}_k \bm{\Lambda}_t (\bm{\mu} + \bm{\mathrm{J}} \bm{\Lambda}_t)^{\top}$$
	where $\bm{\mathrm{J}}_k$ is a 2 by $2K$ matrix such that
	$$
	[ \bm{\mathrm{J}}_k]_{ij} = 
	\left\{
	\begin{array}{@{}ll@{}}
		1, & \text{if}\ i=1\ \mathrm{and}\ j=2k-1\ \mathrm{or}\ i=2\ \mathrm{and}\ j=2k\\
		0, & \text{otherwise.}
	\end{array}\right.
	$$
\end{remark}

\begin{proposition}\label{Prop:second}
	Under the steady state condition,
	$$ \E[\bm{N}_t \bm{N}_t^{\top}] = \A t^2 + \left( \B + \B^{\top} + \Dg(\E[\bm{\lambda}_t]) \right) t $$
	where $\A$ and $\B$ are defined by Eqs.~\eqref{Eq:A} and \eqref{Eq:B}, respectively.
\end{proposition}

\begin{proposition}~\label{Prop:var}
	Under simple calculation, we have
	$$\Var(N_1(t) - N_2(t)) = 2 \begin{bmatrix} 1 & -1 \end{bmatrix} \B \begin{bmatrix} 1 \\ -1 \end{bmatrix}
	+ \begin{bmatrix} 1 & 1 \end{bmatrix}  \E[\bm{\lambda}_t] .$$
\end{proposition}

\subsection{Conditional concavity}

Multi-kernel Hawkes models are inferred by the maximum likelihood estimation (MLE).
It is worthwhile to verify whether the numerical procedure of MLE is performed properly
because of the complexity of the model.
We discuss the issue on finding the global maximum of the log-likelihood function
in terms of the conditional concavity introduced by \cite{lee2017marked} in a single-kernel framework.

Consider a general case of the Hawkes model:
$$ 
\bm{N}_t =
\begin{bmatrix} 
N_1(t) \\
\vdots\\
N_m(t) \\
\end{bmatrix}, \quad
\bm{\lambda}_t = 
\begin{bmatrix} 
\lambda_1(t) \\
\vdots\\
\lambda_m(t) \\
\end{bmatrix}
$$ 
where the intensities follow the $m$-dimensional version of Eq.~\eqref{Eq:lambda}.
The component of $\bm{h}$ is 
\begin{equation*}
h_{ij}(t) = \sum_{k=1}^K \alpha_{ijk}\e^{-\beta_{ijk} t}.
\end{equation*}
The log-likelihood function is 
\begin{align}
L(\mu, \alpha, \beta) = \sum_{i=1}^{m} \left(\sum_{n=1}^{N_i(T)} \log\lambda_i (t_{i,n}) - \int_0^T \lambda_i(u) \D u \right)\label{Eq:loglikelihood}
\end{align}
where $\{t_{i,n}\}$ are the $n$-th jump times of $N_i$.
We rewrite
\begin{align}
\lambda_i(u) &= \mu_i +
\sum_{j=1}^{m}\sum_{k=1}^{K}\sum_{n=1}^{N_j(T)} \alpha_{ijk} \e^{-\beta_{ijk} (u-t_{j,n})}\mathbbm{1}_{\{ u>t_{j,n}\}}  + \varepsilon_i(u)\label{Eq:approxlambda} \\
& \approx \mu_i +
\sum_{j=1}^{m}\sum_{k=1}^{K}\sum_{n=1}^{N_j(T)} \alpha_{ijk} \e^{-\beta_{ijk} (u-t_{j,n})}\mathbbm{1}_{\{ u>t_{j,n}\}} \label{Eq:approxlambda2}
\end{align}
where
\begin{align*}
\varepsilon_i(u) = \sum_{j=1}^{n} \sum_{k=1}^{K} \int_{-\infty}^{0} \alpha_{ijk} \e^{-\beta_{ijk} (u-s)} \D N_j(s)
\end{align*}
which is the remaining impact on $\lambda_i$ by the events which occur before time zero
and this term can be ignored.
If $\beta_i := \beta_{ijk}$ for all $j$ and $k$, then,
$$ \varepsilon_i(u) = (\lambda_i(0) - \mu_i) \e^{-\beta_i u}.$$
If the system is Markovian,
then the assumption that $\varepsilon$ can be ignored is unnecessary.

By integrating Eq.~\eqref{Eq:approxlambda2},
\begin{align*}
\int_0^T \lambda_i(u) \D u  &\approx \mu_i T 
+ \sum_{j=1}^{m}\sum_{k=1}^{K}\sum_{n=1}^{N_j(T)} b_{ijk, n}\alpha_{ijk}, \quad b_{ijk, n} := \frac{1 -\e^{-\beta_{ijk}(T-t_{j,n})}}{\beta_{ijk}},
\end{align*}
and by substituting the above in Eq.~\eqref{Eq:loglikelihood},
\begin{align*}
L(\mu,\alpha,\beta) &\approx \sum_{i=1}^{m} \left\{- \mu_i T +
\sum_{n=1}^{N_i(T)}  \left( \log \lambda_i (t_{i,n})
- \sum_{j=1}^{m} \sum_{k=1}^{K} b_{jik,n} \alpha_{jik} \right) \right\} \\
&= \sum_{i=1}^{m} \left\{ - \mu_i T  +
\sum_{n=1}^{N_i(T)}  \left( \log \left(\mu_i +  \sum_{j=1}^{m}\sum_{k=1}^{K}a_{ijk,n} \alpha_{ijk}\right)
- \sum_{j=1}^{m} \sum_{k=1}^{K} b_{jik,n} \alpha_{jik} \right) \right\}
\end{align*}
where
$$ a_{ijk, n} =  \int_{-\infty}^{t_{i,n}}  \e^{-\beta_{ijk} (t_{i,n} - u)} \D N_{j} (u)$$
and we use
\begin{align*}
\lambda_i(t_{i,n}) = \mu_i + \sum_{j=1}^{m}\sum_{k=1}^{K} \int_{-\infty}^{t_{i,n}}  \alpha_{ijk} \e^{-\beta_{ijk} (t_{i,n} - u)} \D N_{j} (u) .
\end{align*}

We examine the conditional concavity of $L(\mu, \alpha | \beta)$ with fixed $\beta$s with the conditional Hessian.
Let
$$ L_{i,n}(\mu, \alpha | \beta) = \log \left(\mu_i +  \sum_{j=1}^{m}\sum_{k=1}^{K}a_{ijk,n} \alpha_{ijk}\right)
- \sum_{j=1}^{m} \sum_{k=1}^{K} b_{jik,n} \alpha_{jik},$$
then
$$ L(\mu, \alpha | \beta) \approx \sum_{i=1}^{m} \left( -\mu_i T + \sum_{n=1}^{N_i (T)} L_{i,n}(\mu, \alpha | \beta) \right).$$
Since the second derivatives of $\mu_i T$ with respect to $\mu_i$s and $\alpha_{ij}$s are all zeros, 
it is sufficient to check the second derivatives of $L_{i,n}(\mu, \alpha | \beta)$ to examine the Hessian of $L(\mu, \alpha | \beta)$.
Note that
\begin{align*}
&\frac{\partial^2 L_{i,n}}{\partial \mu_i^2 } = - \frac{1}{\lambda_i^2(t_{i,n})}, \quad\quad
\frac{\partial^2 L_{i,n}}{\partial \mu_i \partial \alpha_{ijk} } =  - \frac{a_{ijk,n}}{\lambda_i^2(t_{i,n})},
\\
&\frac{\partial^2 L_{i,n}}{\partial \alpha_{ijk} \partial \alpha_{ij'k'}} =
\begin{cases}
- \dfrac{a_{ijk,n} a_{ij'k',n}}{\lambda_i^2(t_{i,n})}, & \text{if } j \neq j' \text{ and }  k \neq k' \\
0, & \text{otherwise.}
\end{cases}
\end{align*}

As the Hessian of $L_{i,n}(\mu, \alpha | \beta)$ is negative semidefinite, 
the Hessian of $L(\mu, \alpha | \beta)$ is and hence $L(\mu, \alpha | \beta)$ is concave.
Due to the conditional concavity, a numerical optimizer can find 
$$ L^{*}(\beta) := \max_{\mu, \alpha} L(\mu, \alpha | \beta).$$
for each $\beta$ (if it exists).
Therefore, once we determine $L^{*}(\beta)$ for enough $\beta$s and
if we can calculate the maximum of $L^{*}(\beta)$, 
it is possible to determine the global maximum of $L(\mu, \alpha, \beta)$.
Given that this strategy takes a long time, it is only used in this study when examining the global maximum of the Hawkes log-likelihood. 
For the empirical studies, 
a typical quasi-Newton method is applied under the assumption that the numerical optimizer works well.

\subsection{Simulation study}

First, by comparing the theoretically calculated moment values derived in Subsection~\ref{Subsect:moment} with the sample mean obtained through simulation, we verify whether the simulation process is performed well.
The simulation method is based on the thinning algorithm, 
which can generally be applied to nonhomogeneous Poisson processes \citep{Lewis1979}.

\begin{example}~\label{Ex1}
	Under the three kernel model, with presumed parameters
	$$\bm{\mu} = \begin{bmatrix} 0.0757 \\ 0.0757 \end{bmatrix}, \enspace
	\bm{\alpha}_1= \begin{bmatrix} 23.34 & 15.67 \\ 15.67 & 23.34 \end{bmatrix}, \enspace
	\bm{\alpha}_2= \begin{bmatrix} 6 & 9 \\ 9 & 6 \end{bmatrix}, \enspace
	\bm{\alpha}_3= \begin{bmatrix} 0.10 & 0.02 \\ 0.02 & 0.1 \end{bmatrix},
	$$
	$$
	\bm{\beta}_1 = \begin{bmatrix} 140 & 0 \\ 0 & 140\end{bmatrix}, \enspace
	\bm{\beta}_2 = \begin{bmatrix} 30 & 0 \\ 0 & 30\end{bmatrix}, \enspace
	\bm{\beta}_3 = \begin{bmatrix} 0.8 & 0 \\ 0 & 0.8\end{bmatrix}, 
	$$
	and $t = 1,000$, by Propositions~\ref{Prop:1} and \ref{Prop:second},
	we have
	$$ \E[\bm{N}_t] = \E[\bm{\lambda}_t]t = \begin{bmatrix} 1,059.8 \\ 1,059.8 \end{bmatrix}, \enspace
	\E[\bm{N}_t \bm{N}_t^{\top}] = \begin{bmatrix} 1,227,649 & 1,226,463 \\ 1,226,463 & 1,227,649 \end{bmatrix}.
	$$
	Our simulation with $10,000$ generated paths shows that the sample means
	$$ 
	\overline{N_{1}(t)} = 1,058, \quad 
	\overline{N_{1}^2(t)} = 1,218,060, \quad 
	\overline{N_{1}(t) N_{2}(t)} = 1,215,964
	$$
	which confirms that the simulation process works well.
\end{example}

Next, we examine the estimation performance of the model.
The concavity of the log-likelihood depends on the sample size and the branching ratio.
If the number of observations is insufficient and the branching ratio or $\alpha$ is close to zero, then the log-likelihood function is probably not concave.
The log-likelihood function is likely concave with sufficient sample size and significant $\alpha$,
Note that the Hawkes process with insignificant $\alpha$ compared with $\beta$ is very close to the homogeneous Poisson;
it is evident that the model with insignificant $\alpha$ is not well estimated.
The simulation study shows that the more significant $\alpha$, the lower the sample size that is needed and
if $\alpha$ is close to zero, the estimation could be incorrect.

\begin{figure}
	\centering
	\includegraphics[width=0.45\textwidth]{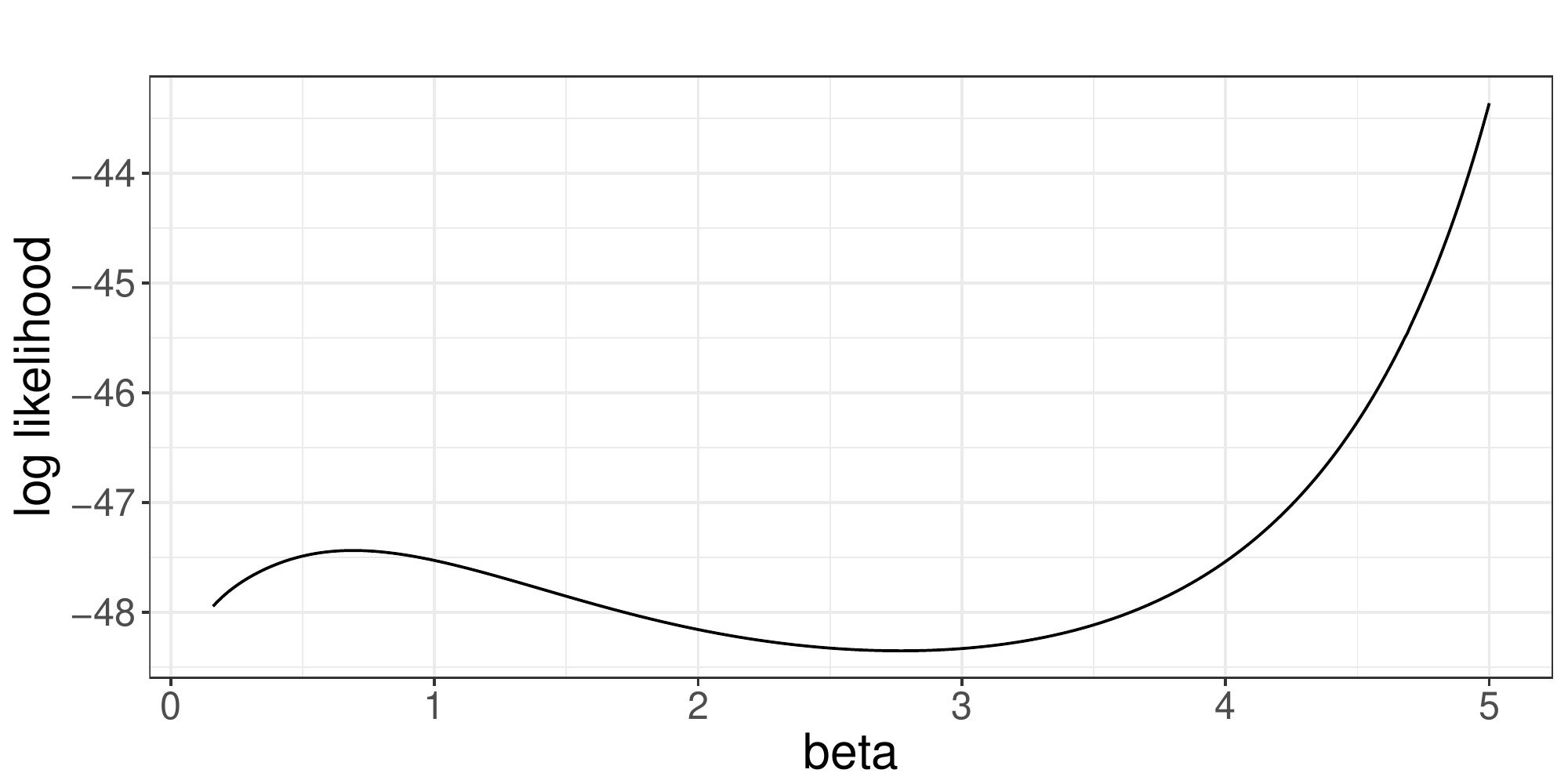}
	\quad
	\includegraphics[width=0.45\textwidth]{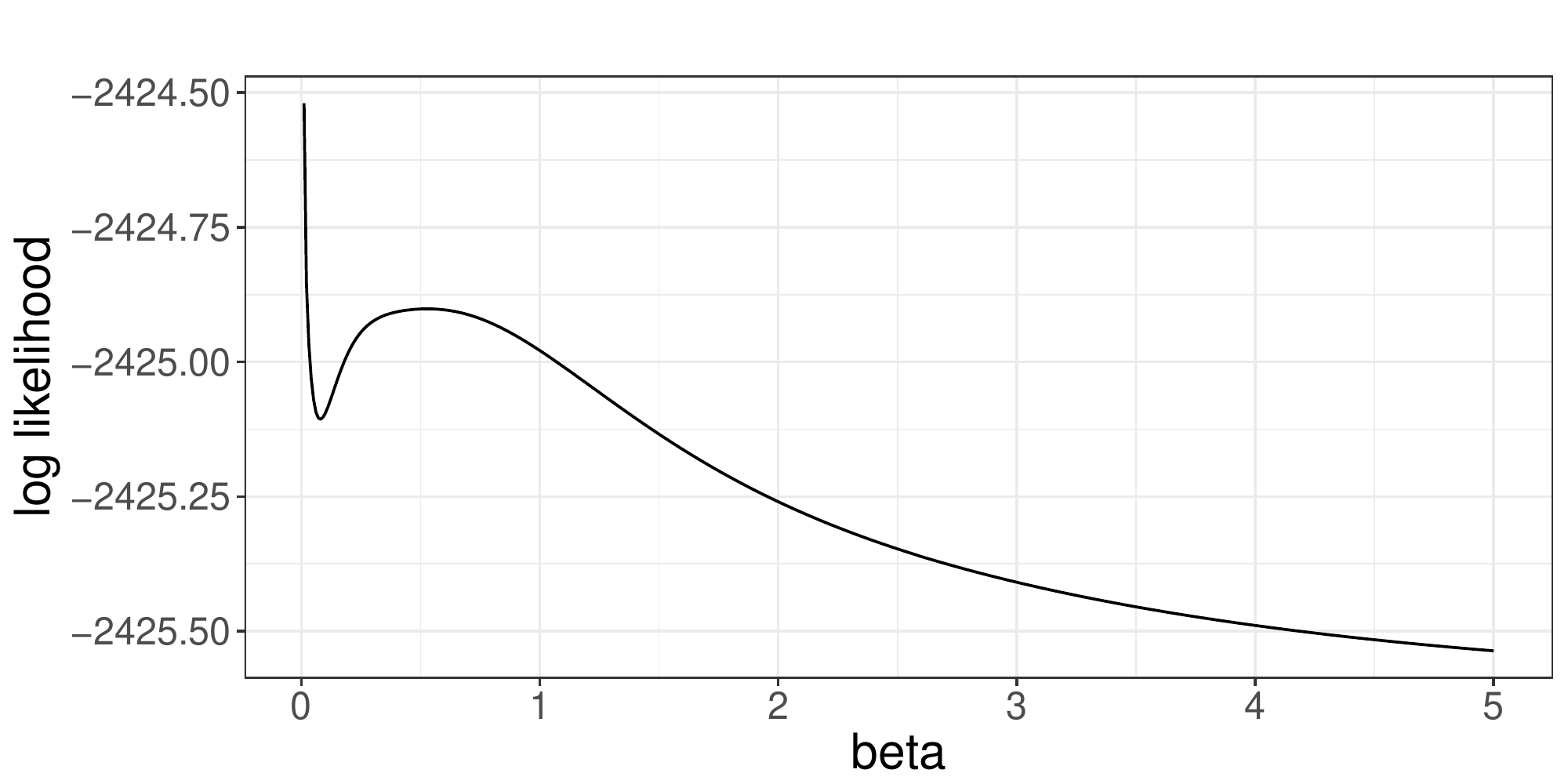}
	\caption{Example of numerically computed conditional maximum of log-likelihood function, $L^{*}$, for each $\beta$ with simulated paths, small sample size case (left) and insignificant $\alpha$ case (right)}\label{Fig:llh}
\end{figure}

Figure~\ref{Fig:llh} illustrates examples of non-concave log-likelihood functions in a univariate Hawkes model.
The left represents the small sample size case with parameters $\alpha = 0.15, \beta = 1.0$ and 10 observations,
while the right represents the case of insignificant $\alpha = 0.001$ with $\beta = 1.0$ and sample size 500.
Using a numerical optimizer, we determine the $L^{*}(\beta)$  for each $\beta$ on a sufficiently dense set.
As non-concavities are observed in the figure, the numerical optimizer may fail to find the global maximum.

A small $\alpha$ value can cause an inaccurate estimation for $\beta$.
In a multivariate Hawkes model, if $\beta$s are considered insignificantly different from each other,
it would be preferable to assume that $\beta$s are equal in the model for parsimony.
For example, consider the following model:
\begin{equation*}
\begin{bmatrix} 
\lambda_1(t) \\
\lambda_2(t) 
\end{bmatrix} = 
\begin{bmatrix} 
0.2 \\
0.2 
\end{bmatrix} + 
\int_{(0,t]}  
\begin{bmatrix} 
0.5\e^{-2.1(t-u)} & 0.001\e^{-2.3(t-u)} \\
0.9\e^{-2.2(t-u)} & 0.5\e^{-2.0(t-u)}
\end{bmatrix}
\, 
\begin{bmatrix} 
\D N_1(u) \\
\D N_2(u) 
\end{bmatrix}
,\label{Eq:intensity}
\end{equation*}
where $\alpha_{12} = 0.001$ is close to zero, and $\beta_{ij}$s differ insignificantly.
With a simulated path with 5,000 observations,
the estimates of $\beta$s are
$$
\begin{bmatrix} 
\hat \beta_{11} & \hat \beta_{12} \\
\hat \beta_{21} & \hat \beta_{22}
\end{bmatrix}
=
\begin{bmatrix} 
2.28 & 0.14 \\
2.33 & 2.35
\end{bmatrix}
$$
where $\hat \beta_{12}$ is inaccurate.
With a constraint of all equal $\beta_{ij}$, the estimate of $\beta_{11}=\beta_{12}=\beta_{21}=\beta_{22}$ is 2.3048, which is much more reliable.

We conduct additional experiments to test the relationship between the sample size and the uniqueness of the local maximum.
For computational complexity, the experiment is based on a one-dimensional Hawkes model.
However, it is important to demonstrate that the sample size is crucial. 

\begin{figure}[!th]
	\centering
	\includegraphics[width=0.7\textwidth]{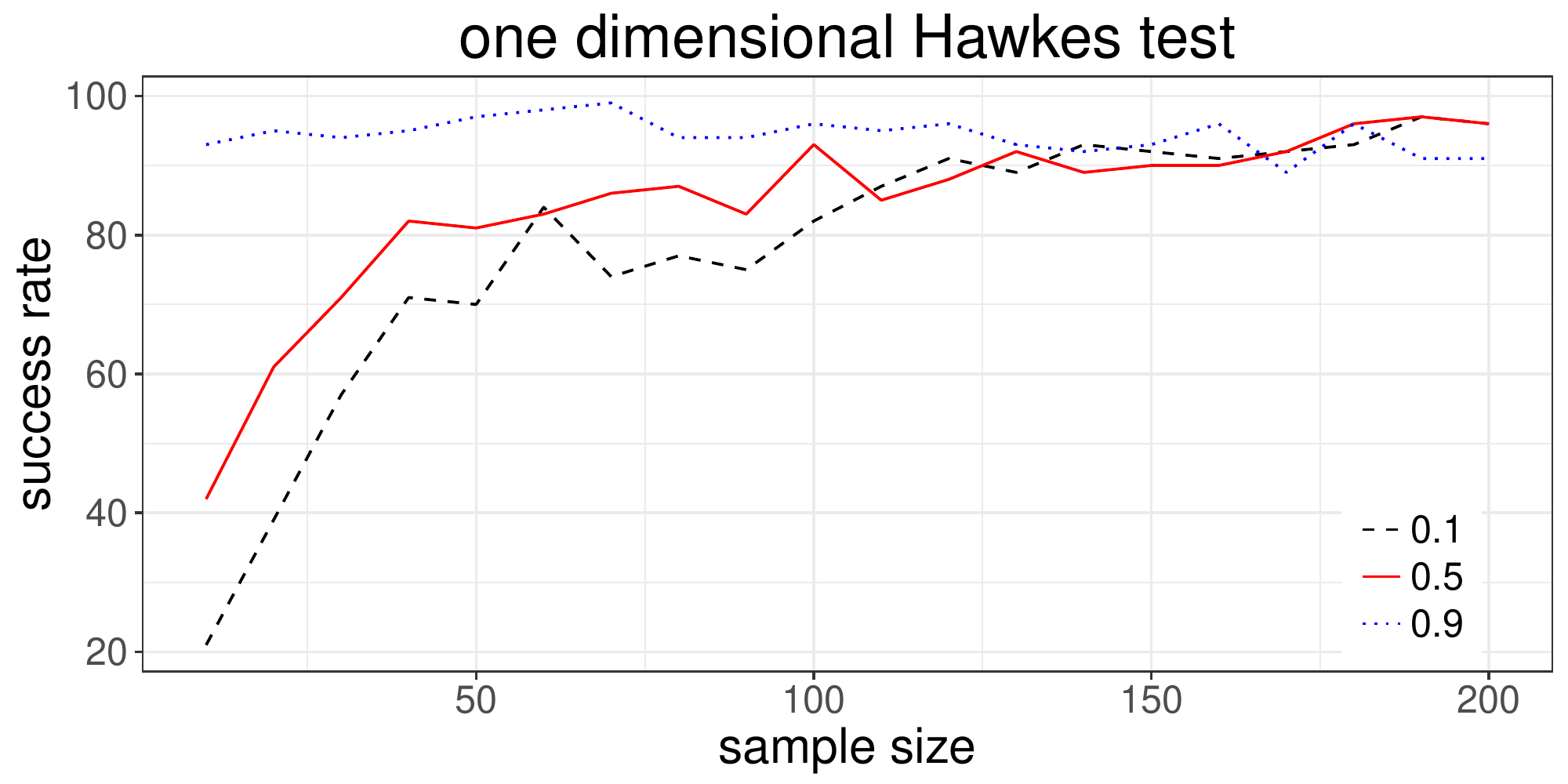}
	\caption{The success rate of estimation to find the global maximum depending on the branching ratio and sample size}\label{Fig:global1}
\end{figure}

A path is simulated for each of the 0.1, 0.5, and 0.9 branching ratios. 
A numerical optimizer finds $L^{*}(\beta)$ for all $\beta$s over a sufficiently dense set.
Examining computed $L^{*}(\beta)$ makes it possible to check the uniqueness of local maximum.
We assume that the numerical optimizer would ultimately succeed in finding the global maximum of $L$
and count it a success, if there is a unique local maximum in $L^{*}$.
Otherwise, we cannot be sure of its success and count this as a failure.
Figure~\ref{Fig:global1} shows success rates after repeating 100 experiments.
As mentioned earlier, the lowest success rate is observed when the branching ratio is low. 
However, with more than 150 samples, the rate is greater than 90\%.
We use more than 10,000 observations in our empirical study.
Furthermore, as the branching ratios are generally significant, we assume that the numerical results are reliable.

\section{Empirical result}\label{sect:result}

\subsection{Basic result}

We conduct empirical studies using high-frequency stock price data under the US equity and futures market from 2016\footnote{From 2016, the data provided by the New York Stock Exchange is recorded in nanoseconds.} to 2019.
The data comprise of the timestamps of changes in the national best bid and offer (NBBO) prices, derived from the trade and quote (TAQ) data released by the New York Stock Exchange.

\begin{figure}[!tb]
	\centering
	\includegraphics[width=0.6\textwidth]{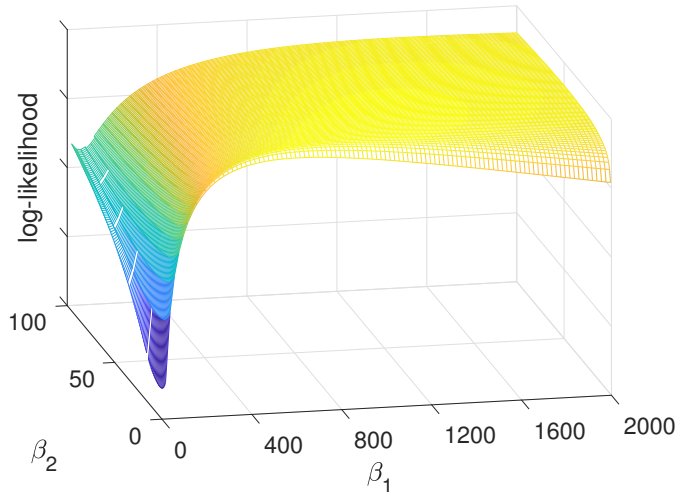}
	\caption{Example of the conditional log-likelihood of the two-kernel Hawkes model as a function of  $\beta$s}\label{Fig:global_two_kernel}
\end{figure}

Before conducting the main analysis, we visualize $L^{*}(\beta_1, \beta_2)$ under the two-kernel model for IBM on January 3, 2017 as shown in Figure~\ref{Fig:global_two_kernel}.
The global maximum is located at $\beta_1 = 880$ and $\beta_2 = 12$.
The other estimates at the maximum are $\hat \mu_1 = \hat \mu_2 = 0.089$, $\hat \alpha_{1s} = 258.8$, $\hat \alpha_{1c} = 102.6$, $\hat \alpha_{2s} = 0.926 $ and $\hat \alpha_{2c} = 1.314$.
The surface is not entirely concave (see the region where $\beta_1$ is close to zero in the figure). 
However, the function is concave and regular around the global maximum.

\begin{figure}[!bt]
	\centering
	\includegraphics[width=0.55\textwidth]{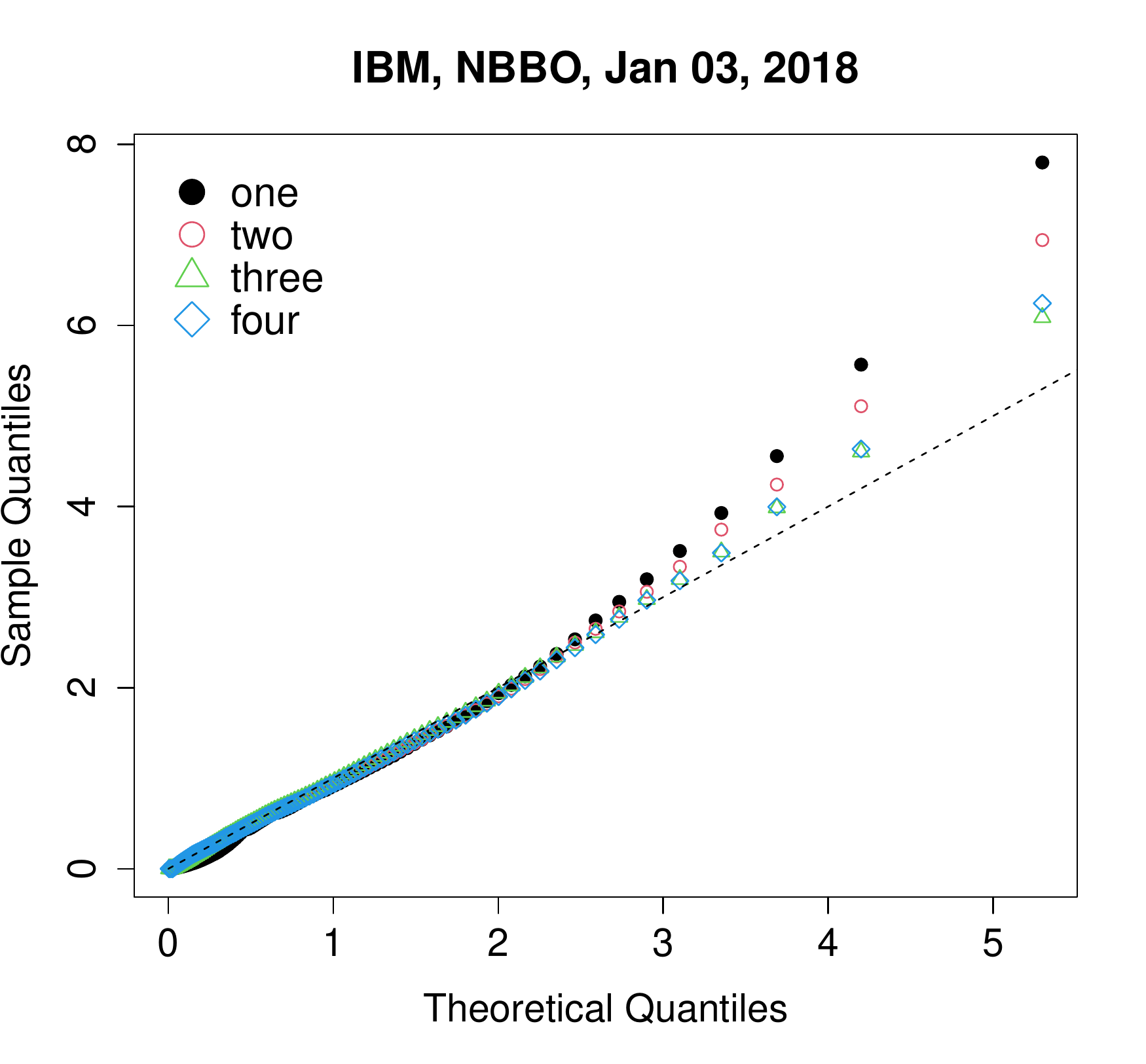}
	\caption{Q-Q plot of residual process of Hawkes models with various kernels}\label{Fig:qqplot}
\end{figure}

Now, we discuss the model-selection issue.
The MLEs are performed on IBM stock's mid-price process with one-, two-, three-, and four-kernel Hawkes models.
In Figure~\ref{Fig:qqplot}, based on IBM data on Jan 03, 2018
we present a Q-Q (quantile-quantile) plot of the residuals versus the unit exponential distribution.
The set of residuals is defined by
$$R = \bigcup R_i, \quad  R_i = \left\{ \int_{t_{i, j}}^{t_{i, j+1}} \hat \lambda_i (u) \D u \right\}$$
where $\hat \lambda_i$ denote inferred intensities.
The points are closer to the straight line, as the number of kernels increases up to three.
Meanwhile, the three-kernel and the four-kernel models differ insignificantly.

\begin{figure}[!hbt]
	\centering
	\includegraphics[width=0.75\textwidth]{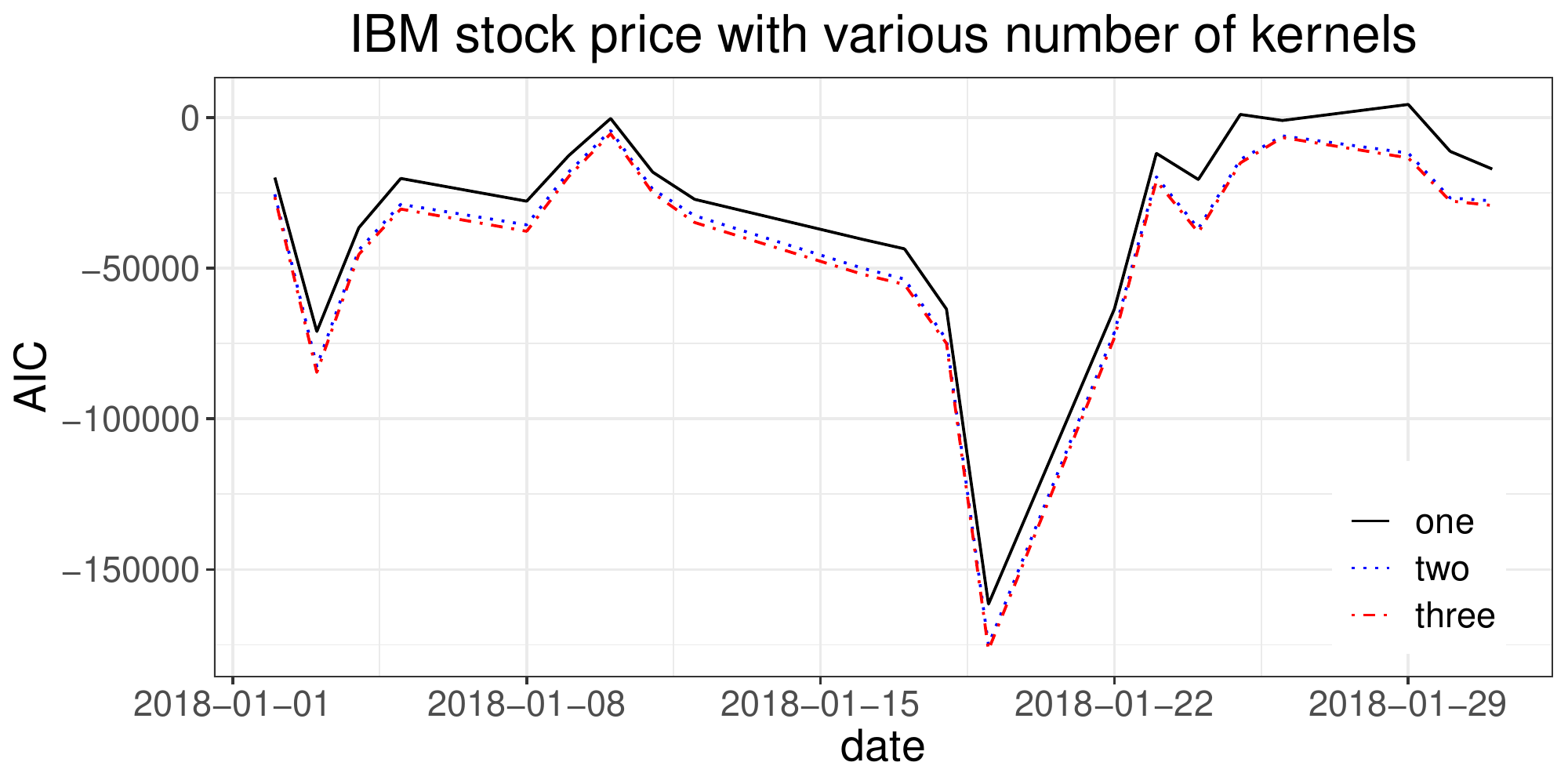}
	\caption{Akaike information criterion (AIC) for one, two and three-kernel models, mid price from NBBO of IBM, January, 2018}\label{Fig:AIC}
\end{figure}

We obtain similar results based on the Akaike information criterion (AIC).
Figure~\ref{Fig:AIC} plots the daily AICs for various kernel models based on IBM data for Jan 2018.
The solid black line, the blue dotted line and the red dash-dot line represent the one-kernel, two-kernel, and three-kernel models, respectively.
The one-kernel model has a significant information loss compared to the others.
The three-kernel model is slightly better than the two-kernel model.
Although not depicted, the four-kernel model has very similar AICs to those of the three-kernel model.
We observe similar patterns in the other cases.
The model extensions are useful for up to two or three kernels but not for higher.

The estimates with a one second time unit under the two-kernel model are reported in Table~\ref{Table:4kernel},
with standard errors in parentheses.
The estimates of $\alpha$s of the first kernel are several hundred;
$\hat \beta$s range from approximately 1,000 to 2,000, 
$\hat \alpha$s of the second kernel are less than 10, and $\hat \beta$s range from approximately 30 to 80.

\begin{figure}[!bt]
	\begin{subfigure}{\textwidth}
		\centering
		\includegraphics[width=0.45\textwidth]{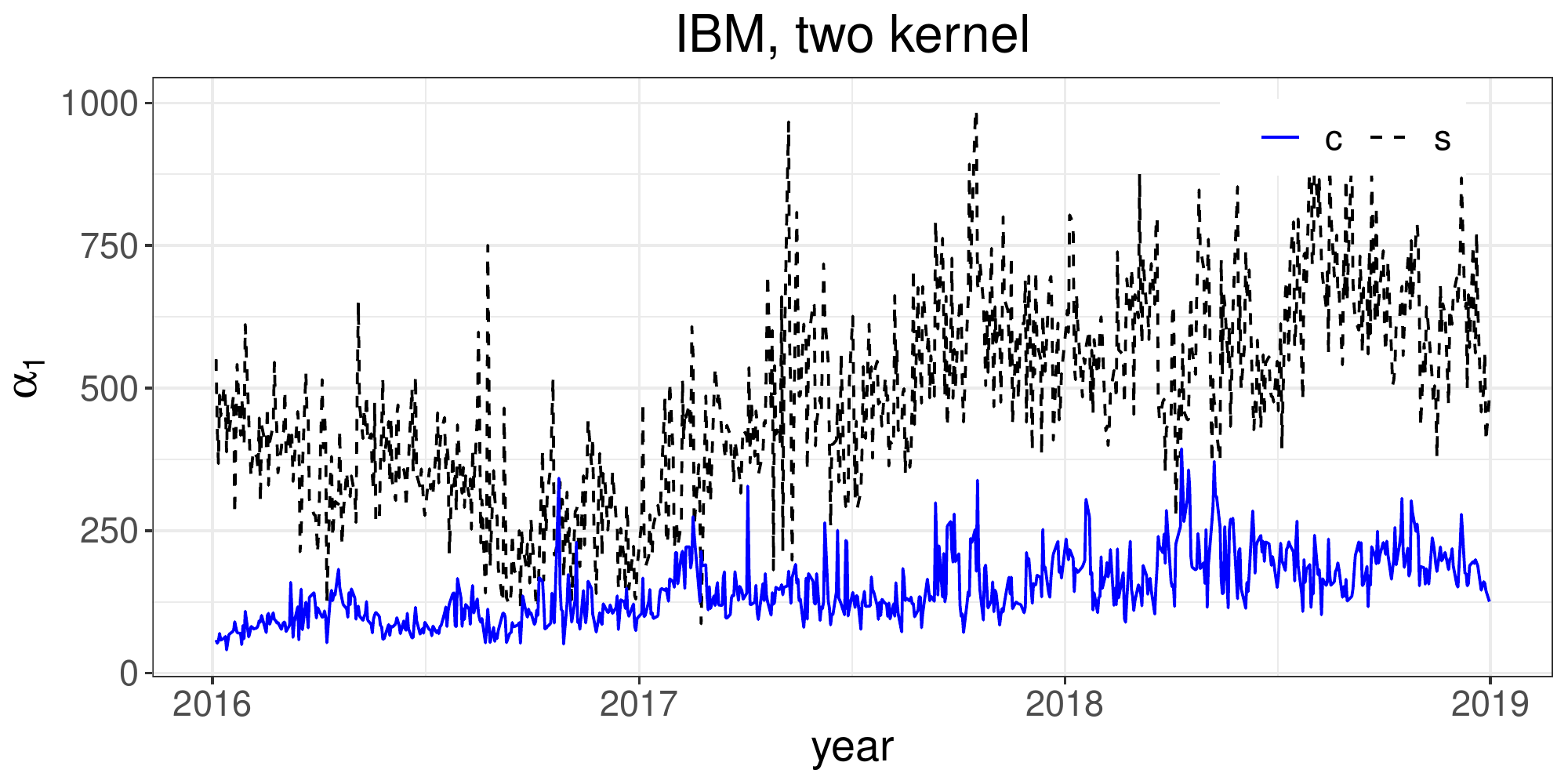}\quad
		\includegraphics[width=0.45\textwidth]{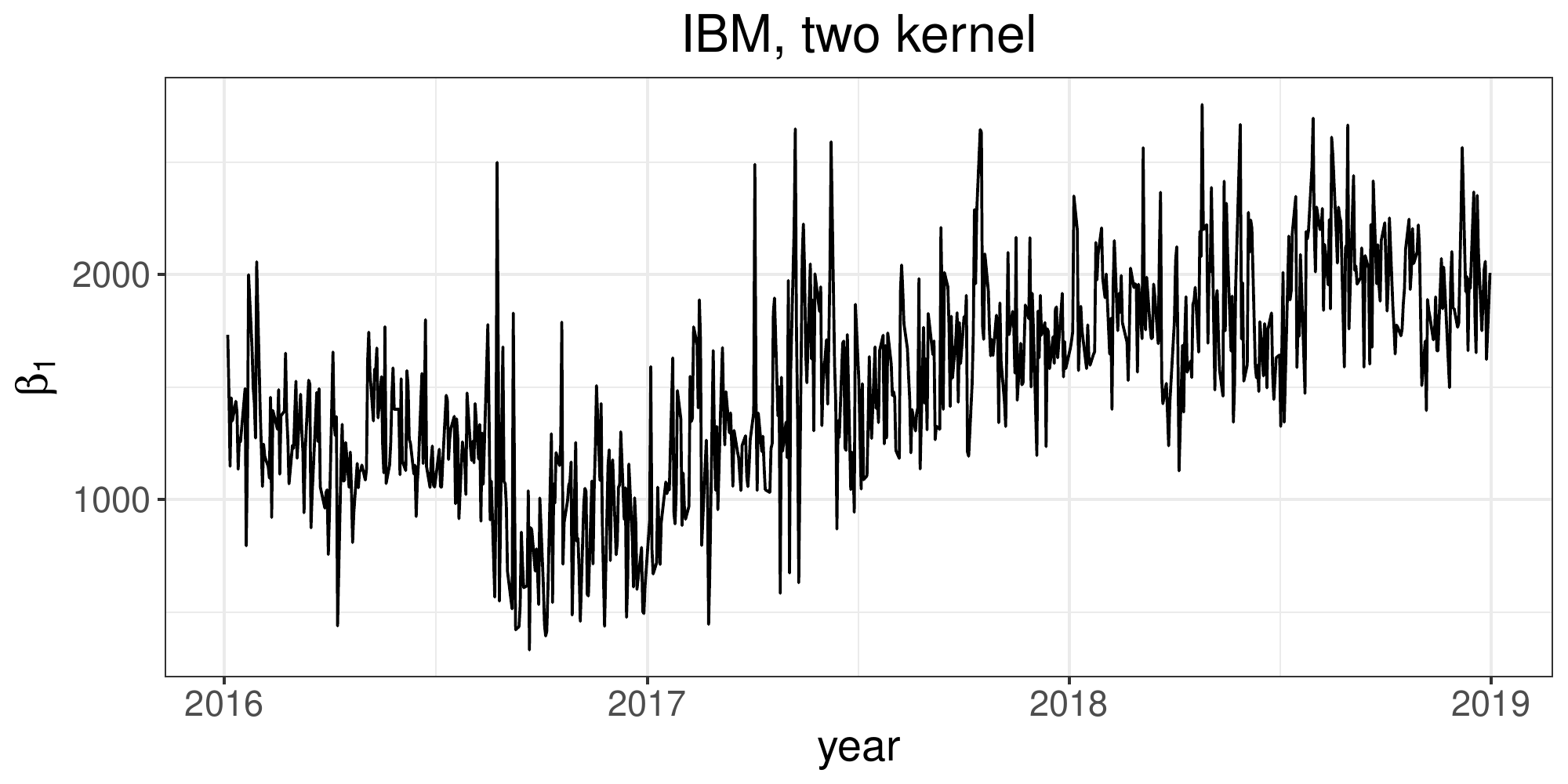}
		\caption{The estimates from ultra-high-frequency kernel, $\alpha_s$ and $\alpha_c$ (left) and $\beta$ (right)}
		\label{fig:uh_two}
	\end{subfigure}
	
	\vspace*{2mm}
	
	\begin{subfigure}{\textwidth}
		\centering
		\includegraphics[width=0.45\textwidth]{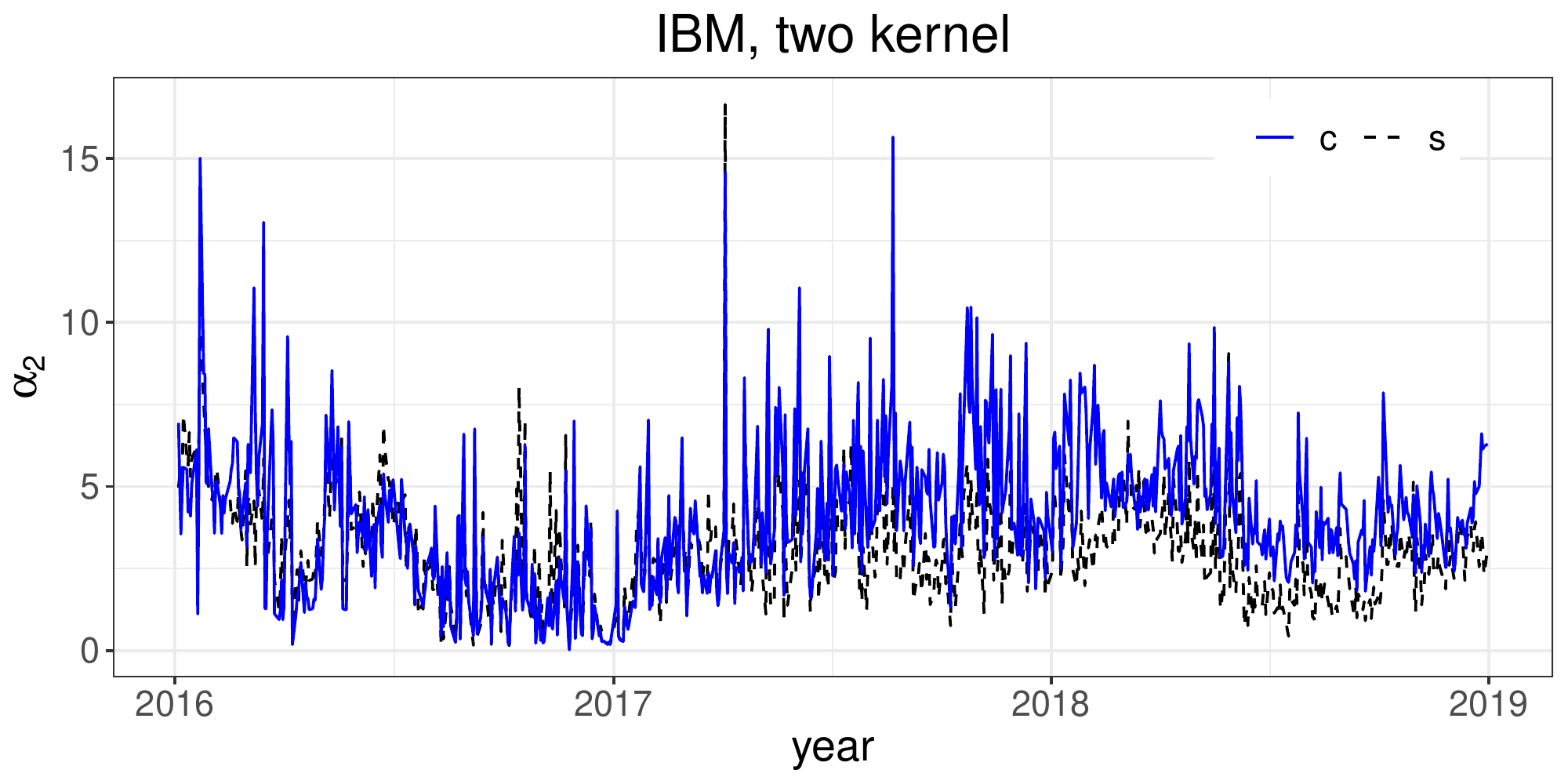}\quad
		\includegraphics[width=0.45\textwidth]{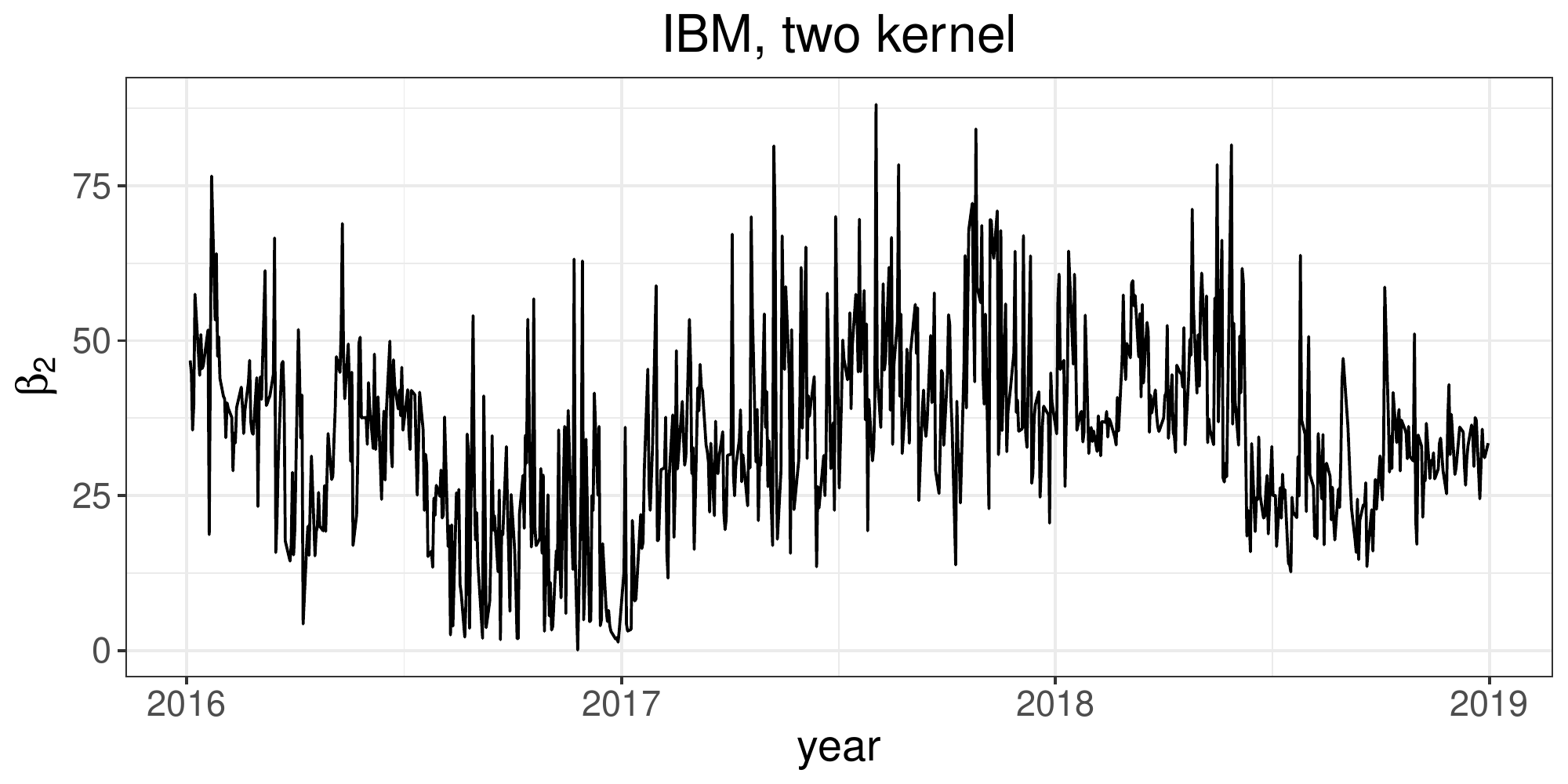}
		\caption{The estimates from high-frequency kernel, $\alpha_s$ and $\alpha_c$ (left) and $\beta$ (right)}
		\label{fig:h_two}
	\end{subfigure}
	\caption{The dynamics of daily estimates of $\alpha$s and $\beta$s under the two-kernel model for IBM stock price}
	\label{Fig:IBM_two_kernel}
\end{figure}

\begin{figure}[!bt]
	\begin{subfigure}{\textwidth}
		\centering
		\includegraphics[width=0.45\textwidth]{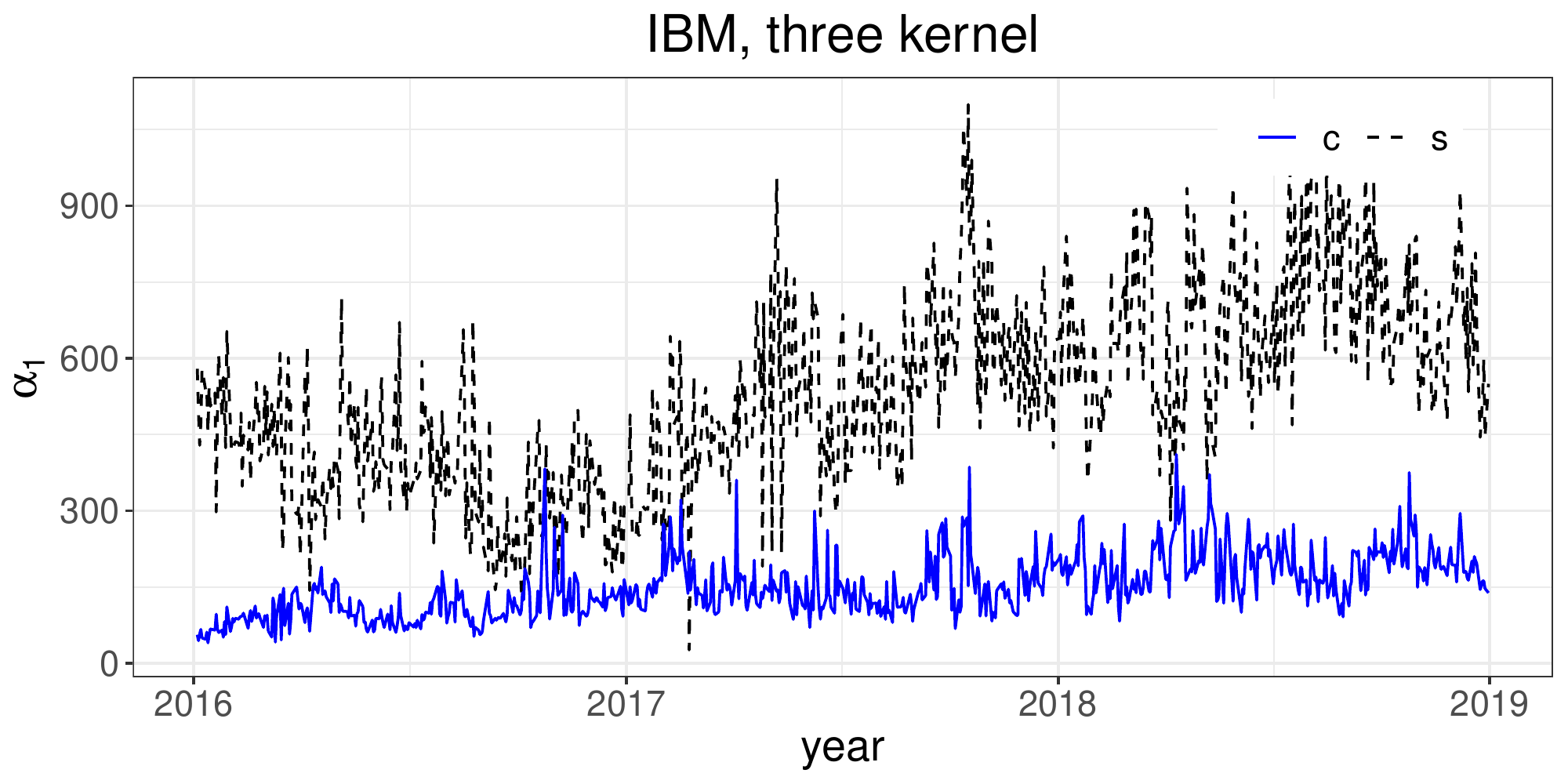}\quad
		\includegraphics[width=0.45\textwidth]{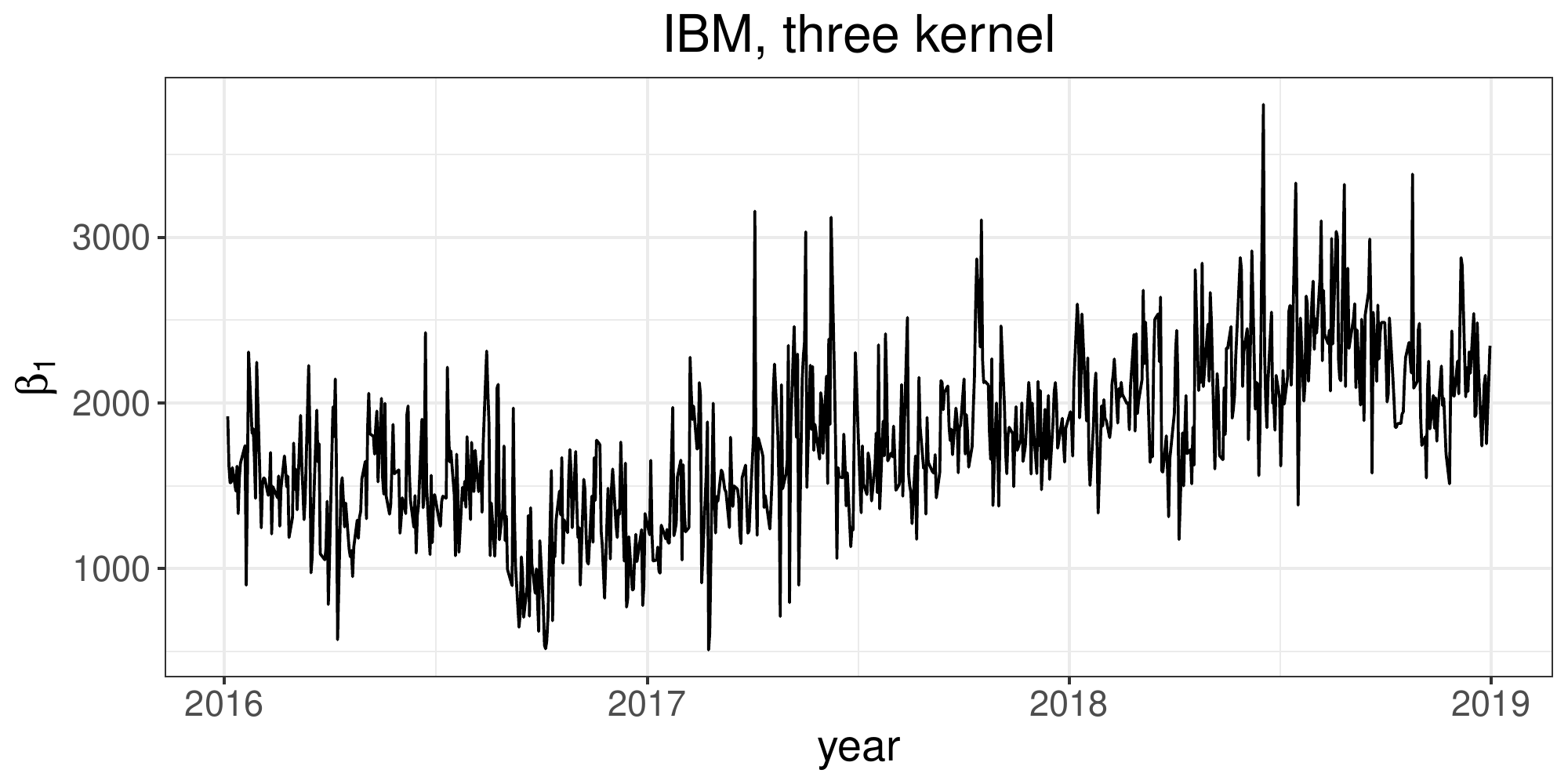}
		\caption{The estimates from ultra-high-frequency kernel, $\alpha_s$ and $\alpha_c$ (left) and $\beta$ (right)}
		\label{fig:uh_three}
	\end{subfigure}
	
	\vspace*{2mm}
	
	\begin{subfigure}{\textwidth}
		\centering
		\includegraphics[width=0.45\textwidth]{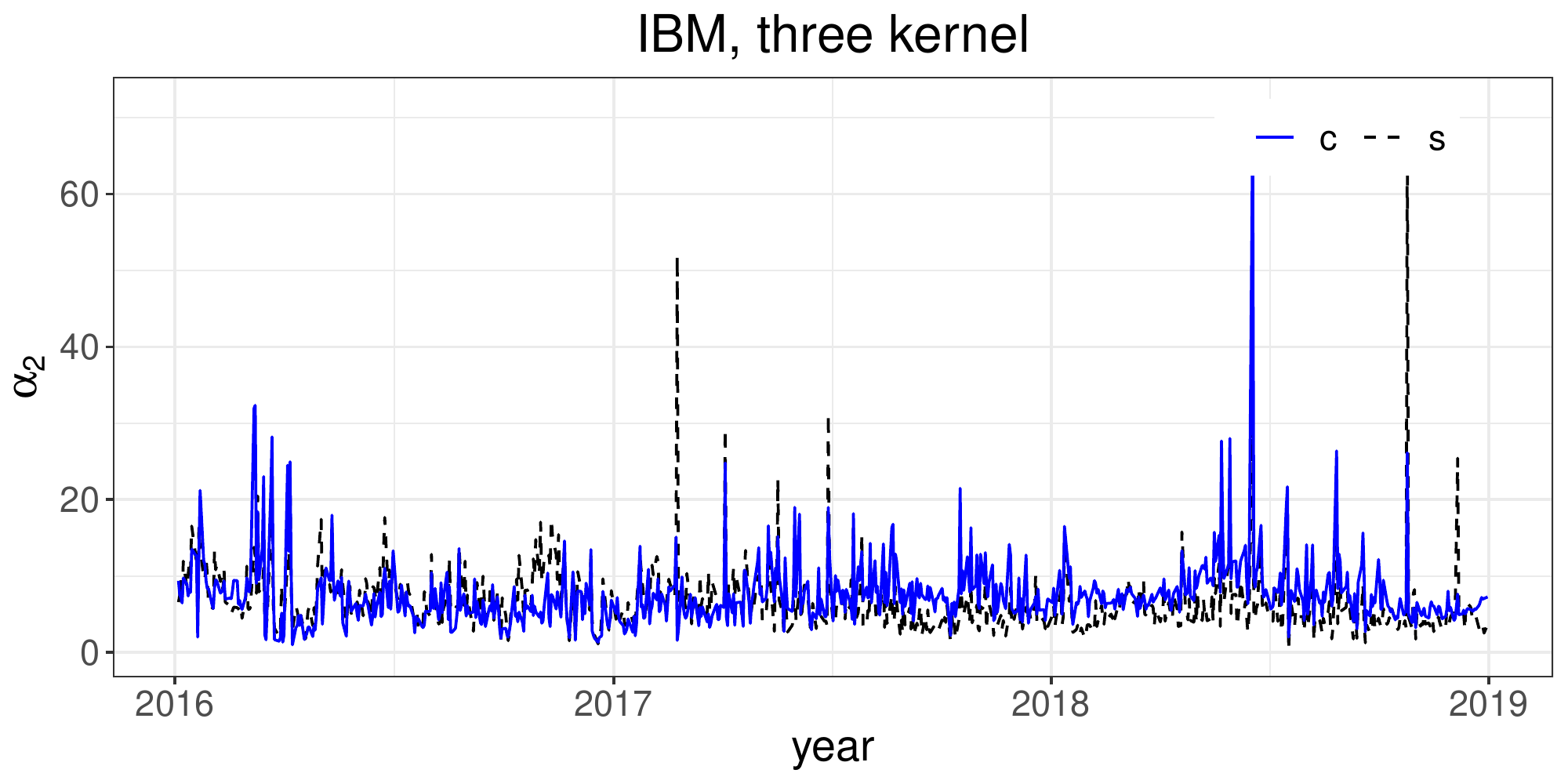}\quad
		\includegraphics[width=0.45\textwidth]{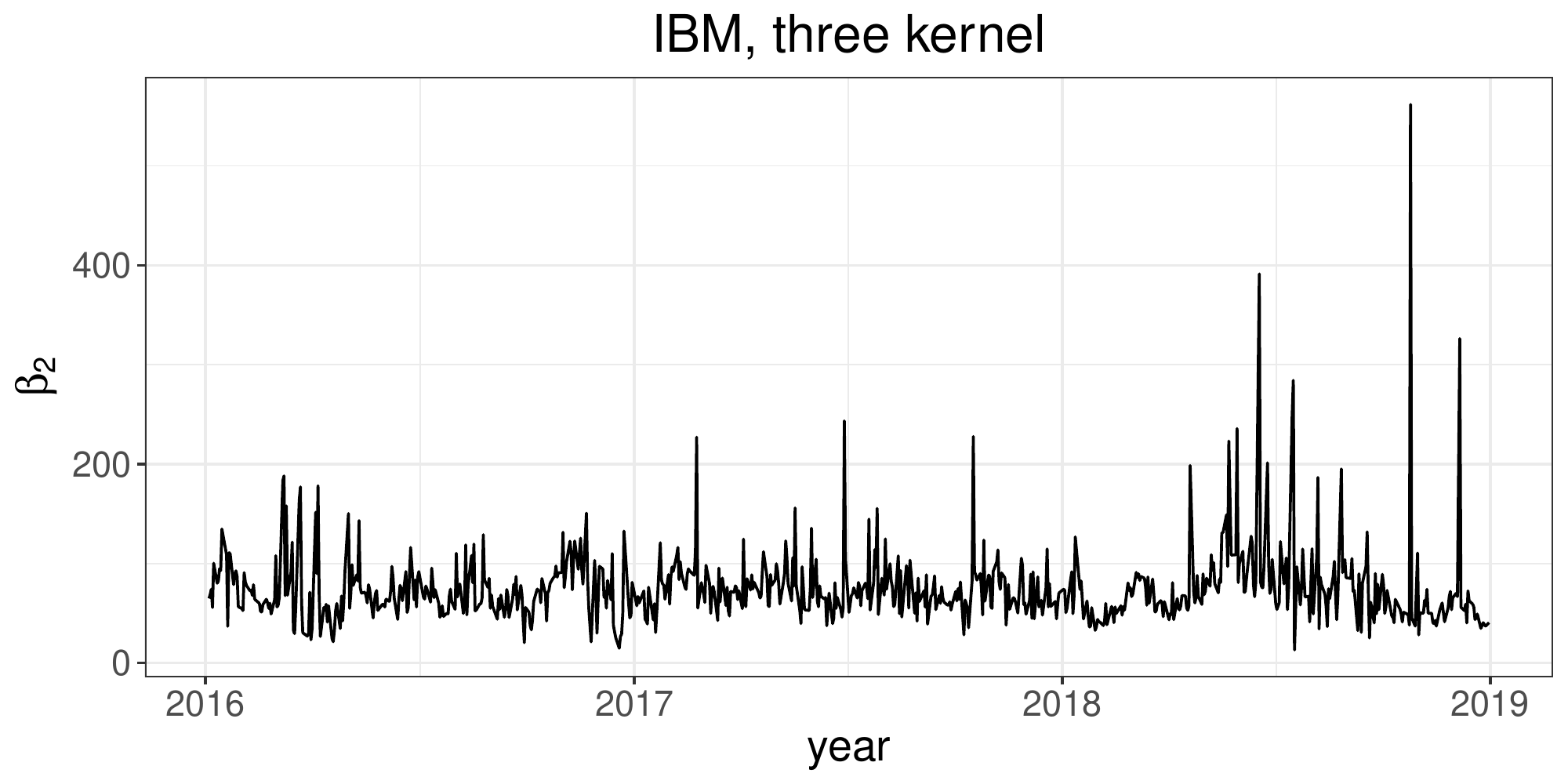}
		\caption{The estimates from very-high-frequency kernel, $\alpha_s$ and $\alpha_c$ (left) and $\beta$ (right)}
		\label{fig:vh_three}
	\end{subfigure}
	
	\vspace*{2mm}
	
	\begin{subfigure}{\textwidth}
		\centering
		\includegraphics[width=0.45\textwidth]{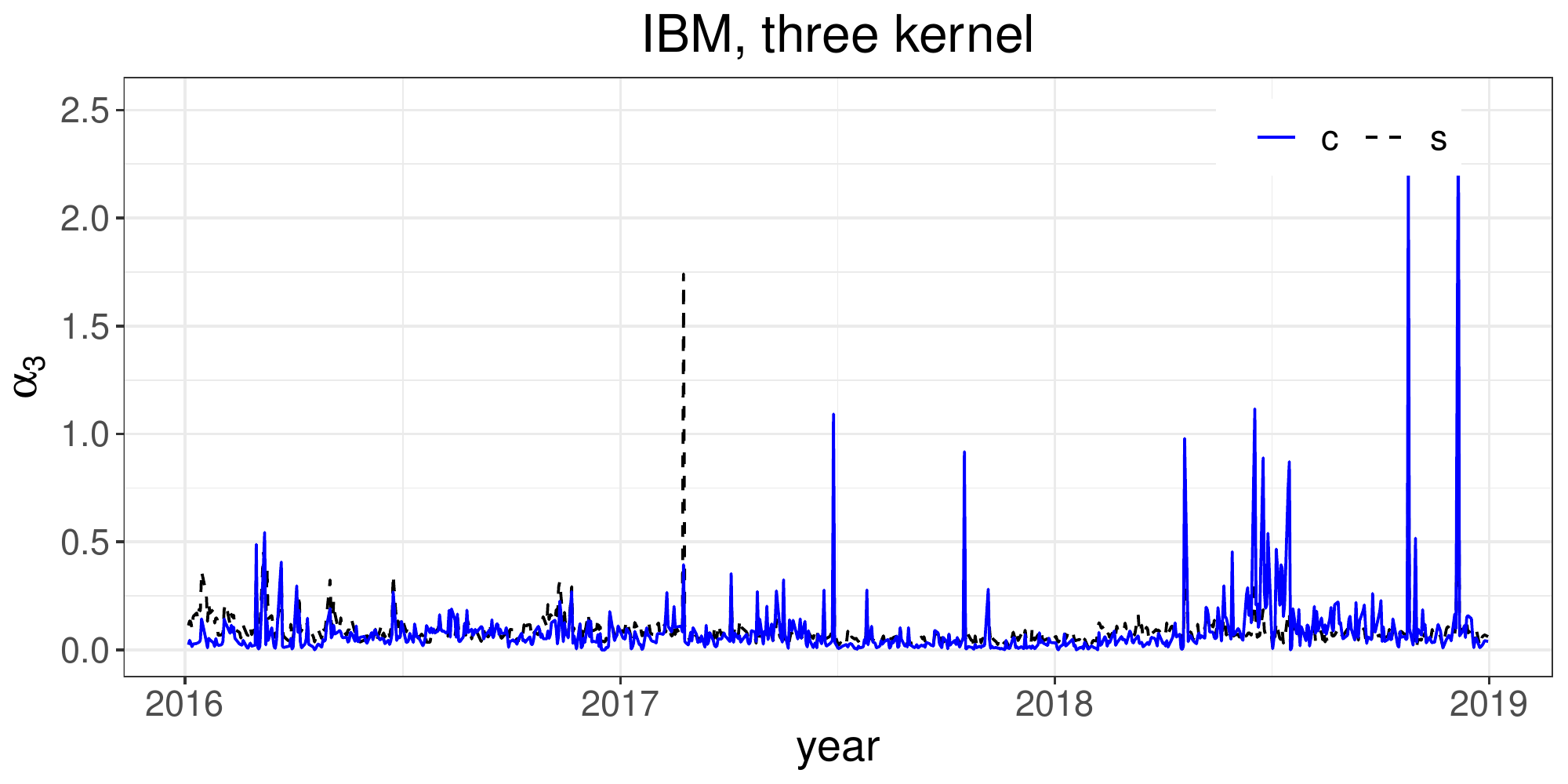}\quad
		\includegraphics[width=0.45\textwidth]{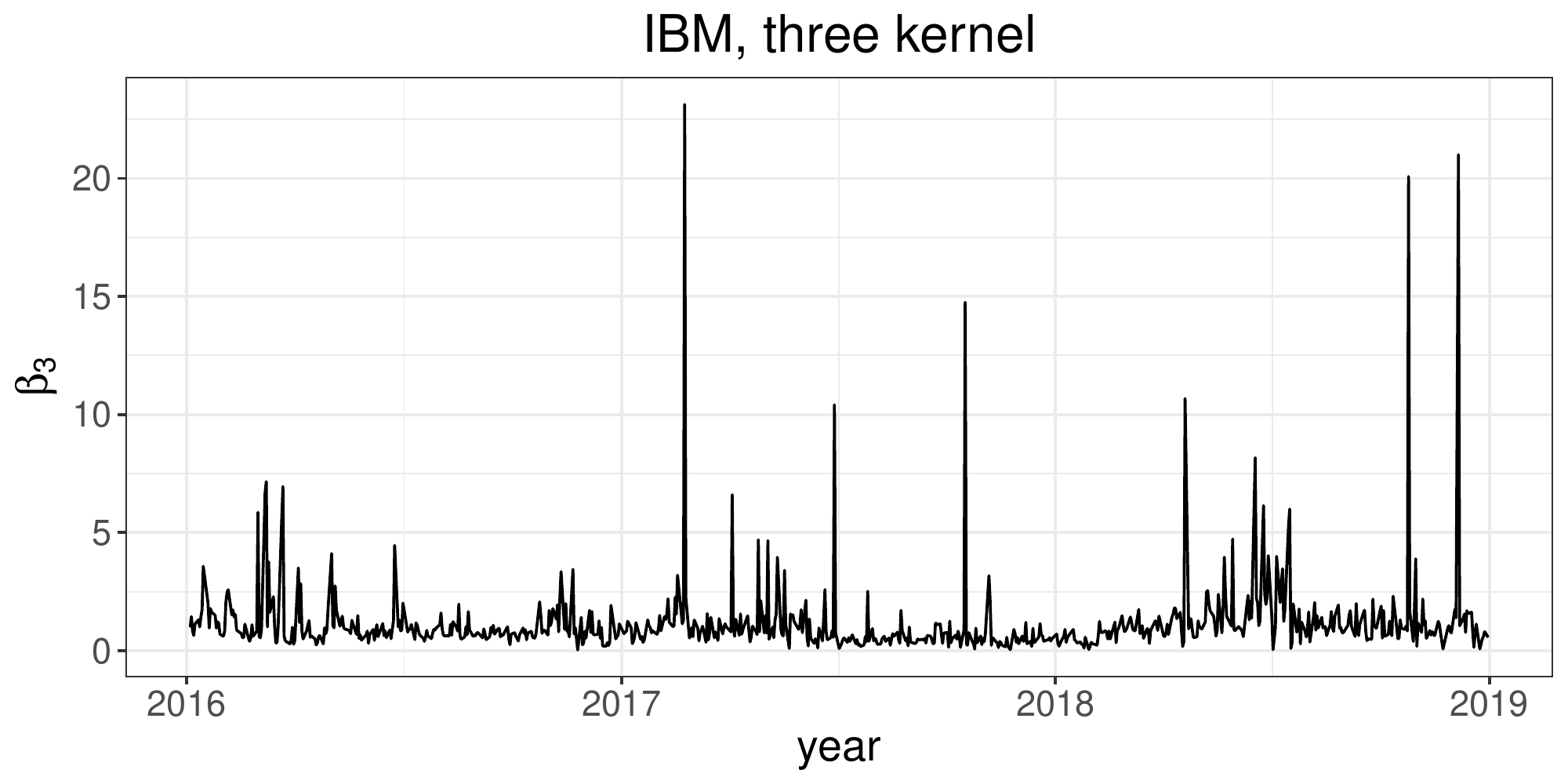}
		\caption{The estimates from the high-frequency kernel, $\alpha_s$ and $\alpha_c$(left) and $\beta$ (right)}
		\label{fig:h_three}
	\end{subfigure}
	
	\caption{The dynamics of daily estimates of $\alpha$s and $\beta$s under the three-kernel model for IBM stock price}
	\label{Fig:IBM_three_kernel}
\end{figure}

Figure~\ref{Fig:IBM_two_kernel}~and~\ref{Fig:IBM_three_kernel} illustrate the dynamics of the daily estimates under the two-kernel and three-kernel models, respectively.
We use the NBBO mid-price processes of IBM from 2016 to 2018.
We term each kernel of the two-kernel model as ultra-high-frequency (UHF) and very-high-frequency (VHF) kernels.
Figure~\ref{fig:uh_two} reveals  $\hat \alpha_{1s}$, $\hat \alpha_{1c}$ and $\hat \beta_1$ for the UHF
and $\hat \alpha_{1s}$s range from approximately 100 to 200 
whereas $\hat \alpha_{1c}$s range from approximately 200 to 1,000.

This implies that, if the mid-price changes, 
the intensities increase instantly at several hundred.
This can lead to numerous additional activities at frequencies of hundreds per second.
Moreover, as shown in the figure, $\hat \beta_1$ is quite large and the effect dissipates quickly.
This fast responsiveness is possibly due to automated high-frequency trading and quotes.
Meanwhile, both $\hat \alpha_{2s}$ and $\hat \alpha_{2c}$ for the VHF kernel ranged from approximately 0 to 15, which is much less than $\hat \alpha$s in the UHF kernel.
Furthermore, $\hat \beta_2$ is distributed between 0 and 80.
This implies a longer persistence than in the UHF kernel.

For the three-kernel model,
we term each kernel to be ultra-high-frequency (UHF), very-high-frequency (VHF), and high-frequency (HF).
The $\hat \alpha$s in the UHF kernel are hundreds,
$\hat \alpha$s in the VHF kernel are up to 60
and $\hat \alpha$s in the HF kernel are less than 2.5.
Table~\ref{Table:summary} presents the summary statistics of the parameter estimates for various models.

\begin{table}[!bt]
	\caption{Summary statistics for estimates from the NBBO of IBM, 2018 under the one, two, and three-kernel Hawkes model for top to bottom}
	\begin{center}
		\begin{tabular}{ccc|ccc|ccc|ccc}
			\hline
			& & & \multicolumn{3}{c|}{UHF} & & & & & & \\
			stock & statistic & $\mu$ & $\alpha_{s}$ & $\alpha_{c}$ & $\beta$ & & & & & & \\
			\hline
			& mean & 0.2768  & 322.8 & 128.6 & 883.5 & & & & & & \\
			IBM & median & 0.2239 & 318.2 & 123.1 & 871.8 & & & & & & \\
			& SD & 0.1595 & 79.86 & 34.68 & 191.4  & & & & & & \\
			\hline
			
			\hline
			& & & \multicolumn{3}{c|}{UHF} & \multicolumn{3}{c|}{VHF} & & & \\
			stock & statistic &  $\mu$ & $\alpha_{s}$ & $\alpha_{c}$ & $\beta$ & $\alpha_{s}$ & $\alpha_{c}$ & $\beta$ & & & \\
			\hline
			& mean & 0.2029 & 615.2 & 193.5 & 1915 & 2.909 & 4.605 & 35.75 & & & \\
			IBM & median & 0.1728 & 619.8 & 188.2 & 1922 & 2.786 & 4.344 & 34.47 & & & \\
			& SD & 0.1055 & 124.5 & 51.69 & 294.8 & 1.376 & 1.619 & 12.63 & & &  \\
			\hline
			
			\hline
			& & & \multicolumn{3}{c|}{UHF} & \multicolumn{3}{c|}{VHF} & \multicolumn{3}{c}{HF}\\
			stock & statistic &  $\mu$ & $\alpha_{s}$ & $\alpha_{c}$ & $\beta$ & $\alpha_{s}$ & $\alpha_{c}$ & $\beta$ & $\alpha_{s}$ & $\alpha_{c}$ & $\beta$\\
			\hline
			& mean & 0.1382 & 677.8 & 196.6 & 2191 & 6.010 & 8.397 & 78.13 & 0.0940 & 0.1271 & 1.449 \\
			IBM & median & 0.1190 & 662.4 & 192.2 & 2132 & 4.862 & 7.457 & 66.77 & 0.0803 & 0.0661 & 1.180 \\
			& SD & 0.0755 & 146.3 & 55.16 & 388.1 & 5.502 & 5.398 & 54.32 & 0.1065 & 0.2501 & 2.055 \\
			\hline
		\end{tabular}
	\end{center}
	\label{Table:summary}
\end{table}

\subsection{Responsiveness}

Under the Hawkes model, the intensities are instantly excited when an event occurs
and decay to the level they would have been at if the excitement had not occurred.
The increased intensity may or may not cause additional event during this period.
If the excitement is large, and the decay is slow, 
then the occurrence of an event is highly possible.
Based on this argument, 
we compute the conditional expectation of the arrival time of an event caused by a single excitement.
This analysis helps us understand the responsive speed of each kernel.

If the excitation parameter is $\alpha$ and the exponential decaying parameter is $\beta$, 
then the remaining amount of excitement in the intensity owing to $\alpha$ when time $s$ has elapsed since the event occurred is
$ \alpha \exp ( -\beta s).$
Based on the survival analysis with a hazard function $\alpha \exp ( -\beta s)$,
the conditional probability for the inter-arrival time, $\tau$, of the next event owing to excitement $\alpha$ is 
\begin{equation}
\PP \{\tau < u < \infty \}  = 1 - \exp \left(- \int_0^u \lambda(s) \D s \right)  = 1 -\exp\left( -\alpha \frac{1 - \e^{-\beta u}}{\beta} \right) \label{Eq:prob_tau}
\end{equation}
Further, the conditional expectation of the next event owing to excitement $\alpha$ is
\begin{equation}
\E[\tau | \tau < \infty] = \int_{0}^{\infty} \alpha u \exp\left( - \alpha \frac{1-\e^{-\beta u}}{\beta} - \beta u \right) \D u \label{Eq:E_tau}
\end{equation}
which can be computed numerically.
Using the above formula and the median estimates of $\alpha$s and $\beta$s for IBM in 2018,
Table~\ref{Table:duration} presents the estimated conditional expectation of the arrival time for each kernel.

\begin{table}[!bt]
	\caption{Conditional expected time of the next arrival by each kernel, IBM, 2018}
	\begin{center}
		\begin{tabular}{c|cc|cc|cc}
			\hline
			kernel & \multicolumn{2}{c|}{UHF} & & & & \\
			\hline
			\multirow{2}{*}{one} & self & cross & & & & \\
			& $319.4$\si{\us} & $145.7$\si{\us} & & & & \\
			\hline
			\hline
			& \multicolumn{2}{c|}{UHF} & \multicolumn{2}{c|}{VHF} & & \\
			\hline
			\multirow{2}{*}{two} & self & cross & self & cross & & \\
			& $132.1$\si{\us} & $59.09$\si{\us} & $2.207$\si{\ms} & $3.327$\si{\ms} & & \\
			\hline
			\hline
		    & \multicolumn{2}{c|}{UHF} & \multicolumn{2}{c|}{VHF} & \multicolumn{2}{c}{HF}\\
		    \hline
		    \multirow{2}{*}{three} & self & cross & self & cross & self & cross \\
		    & $115.8$\si{\us} & $55.61$\si{\us} & $1.032$\si{\ms} & $1.530$\si{\ms} &  $72.74$\si{\ms} & $62.12$\si{\ms} \\
			\hline
		\hline
		\end{tabular}
	\end{center}
	\label{Table:duration}
\end{table}

In the three-kernel model, 
the conditional expectation of the arrival time originating from the UHF kernel ranges from approximately 50 to 100\si{\us}.
The response time is very fast, and the corresponding activities may include automated actions such as flickering quotes and fleeting orders \citep{hasbrouck2009technology}.
The conditional expectation of arrival time by the VHF kernel ranges from approximately 1-2\si{\ms},
while that for the HF kernel, ranges from approximately 60-70 \si{\ms}.
The expected arrival time by the base intensity is approximately 100-200\si{ms}.

Next, we examine the proportions of the causes of events among the kernels and the base intensity.
According to the immigrant offspring argument, 
an event can be an external immigrant represented by the base intensity,
or the descendant of an internally generated offspring described by the kernels.
Based on the inferred intensities derived from the maximum likelihood estimates,
we compute the probability that the parent of an event belongs to the source, that is, kernels or the base intensity.
By comparing the estimated kernel components of the intensities at the time of an event,
we calculate the empirical probability of the causes as the proportion of the intensity component.
More precisely, in the three-kernel model, assuming that an upward move occurs at time $t$, the following 
$$ \frac{\hat \mu_1}{\hat \lambda_1(t)}, \enspace \frac{\hat \lambda_{11}(t)}{\hat \lambda_{1}(t)}, \enspace
\frac{\hat \lambda_{21}(t)}{\hat \lambda_{1}(t)}, \enspace \frac{\hat \lambda_{31}(t)}{\hat \lambda_{1}(t)}$$
represent the estimated probabilities that the cause of event belongs to base intensity, UHF, VHF, and UF kernels, respectively.

The annually aggregated empirical probabilities are presented in Table~\ref{Table:ratio} from 2016 to 2019 for IBM and AAPL.
Approximately 40\% of the events are caused by the UHF kernel which is the largest proportion.
The other kernels have similar percentages, ranging from approximately 15-25\%.
The percentages of the base intensity range from approximately 30-40\% and 15-25\% for the two-kernel and three-kernel models, respectively.

\begin{table}[!bt]
	\caption{The proportion of the causes of mid price changing events under two-kernel model and three-kernel model}
	\begin{center}
		\begin{tabular}{cccccc}
			\hline
			\multicolumn{6}{c}{two-kernel model}\\
			\hline
			stock & year & base & UHF & VHF & HF \\
			\hline
			  IBM & 2016 & 38.40\% & 39.81\% & 21.79\% & \\
			 & 2017 & 37.03\% & 43.96\% & 19.01\% & \\
			 & 2018 & 35.52\% & 42.79\% & 21.69\% & \\
			 & 2019 & 37.35\% & 42.41\% & 20.24\% & \\
			\hline
			 AAPL & 2016 & 35.45\% & 40.44\% & 24.11\% & \\
			 & 2017 & 32.39\% & 43.10\% & 24.50\% & \\
			 & 2018 & 31.34\% & 45.91\% & 22.76\% & \\
			 & 2019 & 33.64\% & 43.31\% & 23.04\% & \\
			\hline
			\multicolumn{6}{c}{three-kernel model}\\
			\hline
			 IBM & 2016 & 26.54\% & 35.67\% & 20.29\% & 17.49\% \\
			 & 2017 & 24.54\% & 41.04\% & 18.51\% & 15.91\% \\
			 & 2018 & 24.22\% & 40.81\% & 19.70\% & 15.27\% \\
			 & 2019 & 25.17\% & 40.65\% & 16.91\% & 17.27\% \\
			\hline
			 AAPL & 2016 & 26.24\% & 34.31\% & 16.53\% & 22.92\% \\
			 & 2017 & 18.70\% & 38.28\% & 20.90\% & 22.12\% \\
			 & 2018 & 16.30\% & 43.83\% & 20.61\% & 19.26\% \\
			 & 2019 & 16.12\% & 40.46\% & 20.42\% & 23.01\% \\
			\hline
		\end{tabular}
	\end{center}
	\label{Table:ratio}
\end{table}

\section{Conclusion}~\label{sect:concl}
The multi-kernel Hawkes model is applied to the high-frequency price process
and the basic moment properties are derived.
Furthermore, the suitability of the MLE is examined in terms of conditional concavity.
The empirical results reveal that high-frequency mid-price dynamics derived directly from raw data follow multi-kernel Hawkes processes.
The diagnostics test show that two- or three-kernel model are appropriate for ultra-high-frequency modeling.

In the two-kernel model, the kernel is composed of the ultra-high-frequency (50-100\si{\us} in expected response time) and very-high-frequency (2-3\si{\ms}).
In the three-kernel model, the kernel is composed of the ultra-high-frequency (50-100\si{\us}), very-high-frequency (1-1.5\si{\ms}), and high-frequency (60-70\si{\ms}).
The percentage estimated by the ultra-high-frequency kernel is approximately 40\%, which is the largest proportion in both the two and three-kernel models.
This percentage is approximately 20\% for the very-high-frequency and high-frequency kernels.
The estimated percentages of external events are 15-25\% and 30-40\% for the three-kernel and two-kernel models, respectively.


\section*{Acknowledgements}
This work has supported by the National Research Foundation of Korea(NRF) grant funded by the Korea government(MSIT)(No. NRF-2021R1C1C1007692).

\bibliography{Bib}
\bibliographystyle{chicago}

\newpage

\appendix

\section{Proofs}~\label{Appendix:proof}

\begin{proof}[Proof for Proposition~\ref{Prop:1}]
	Eq.~\eqref{Eq:dlk} implies
	$$  \bm{\lambda}_{k}(t)  - \bm{\lambda}_{k}(0)  = - \int_0^{t} \bm{\beta}_k \bm{\lambda}_{k}(s) \D s + \int_0^t \bm{\alpha}_k \D \bm{N}_s.$$
	Under the steady state condition, by taking expectation to the both sides of the above equation, we have
	$$ \E[\bm{\lambda}_{k}(t)] - \E[\bm{\lambda}_{k}(0)] = - \bm{\beta}_k \E[ \bm{\lambda}_{k}(t)] t +  \bm{\alpha}_k  \E[\bm{\lambda}_t] t $$
	and using $\E[\bm{\lambda}_{k}(t)] = \E[\bm{\lambda}_{k}(0)]$,  
	$$ \E[\bm{\lambda}_{k}(t)] = \bm{\beta}_k^{-1} \bm{\alpha}_k \E[\bm{\lambda}_t] $$
	and substituting the above to
	$$ \E[\bm{\lambda}_t] = \bm{\mu} + \sum_{k=1}^{K}\E[\bm{\lambda}_{k}(t)],$$
	which is derived by taking the expectation on Eq.~\eqref{Eq:multi},
	we have the desired result.
\end{proof}

\begin{proof}[Proof for Propositoin~\ref{Prop:syl}]
	By Lemma~\ref{Lemma:quad},
	$$
	\bm{\beta} \E[  \bm{\Lambda}_t \bm{\Lambda}_{t}^{\top} ]  + \E[\bm{\Lambda}_{t}  \bm{\Lambda}_{t}^{\top}] \bm{\beta} = \E[\bm{\Lambda}_{t}  \bm{\lambda}_{t}^{\top}]\bm{\alpha}^{\top} + \bm{\alpha} \E [\bm{\lambda}_t \bm{\Lambda}_{t}^{\top}]  + \bm{\alpha} \mathrm{Dg}(\E[\bm{\lambda}_t])\bm{\alpha}^{\top}.
	$$
	Since
	$$ \bm{\lambda}_t = \bm{\mu} + \bm{\mathrm{J}} \bm{\Lambda}_t,$$
	we have
	$$
	(\bm{\beta} - \bm{\alpha}\bm{\mathrm{J}})\E[  \bm{\Lambda}_t \bm{\Lambda}_{t}^{\top} ]  + \E[\bm{\Lambda}_{t}  \bm{\Lambda}_{t}^{\top}] (\bm{\beta} - \bm{\mathrm{J}}^{\top} \bm{\alpha}^{\top}) = \E[\bm{\Lambda}_{t} ]\bm{\mu}^{\top}\bm{\alpha}^{\top} + \bm{\alpha}\bm{\mu} \E [ \bm{\Lambda}_{t}^{\top}]  + \bm{\alpha} \mathrm{Dg}(\E[\bm{\lambda}_t])\bm{\alpha}^{\top}.
	$$
\end{proof}

\begin{proof}[Proof for Propositon~\ref{Prop:lN}]
	For simplicity, consider only the particular solution for $\frac{\D \E[\bm{\lambda}_t \bm{N}_t^{\top}]}{\D t}$.
	Assume that
	$$\E[\bm{\lambda}_t \bm{N}_t^{\top}] \approx \A t + \B, \quad \E[\bm{\lambda}_{k}(t) \bm{N}_t^{\top}] \approx \A_{k}t + \B_k. $$
	Since 
	$$ \E[\bm{\lambda}_t \bm{N}_t^{\top}] = \E\left[\left(\bm{\mu} + \sum_{k=1}^{K}\bm{\lambda}_{k}(t)\right) \bm{N}_t^{\top} \right],$$
	we have
	$$ \A t + \B = \left( \bm{\mu} \E[\bm{\lambda}_t^{\top}] + \sum_{k=1}^{K} \A_k \right) t + \sum_{k=1}^{K} \B_k
	$$
	and
	\begin{align}
		\A = \bm{\mu} \E[\bm{\lambda}_t^{\top}]  + \sum_{k=1}^{K} \A_k , \quad \B = \sum_{k=1}^{K} \B_k . \label{Eq:ABk}
	\end{align}
	Note that
	\begin{align*}
		\D (\bm{\lambda}_{k}(t) \bm{N}_t^{\top}) &= \bm{\lambda}_{k}(t) \D \bm{N}_t^{\top} + (\D \bm{\lambda}_{k}(t)) \bm{N}_t^{\top} + \D [\bm{\lambda}_k \bm{N}^{\top}]_t \\
		&= \bm{\lambda}_{k}(t) \D \bm{N}_t^{\top} +  \left( -\bm{\beta}_k \bm{\lambda}_{k}(t) \D t + \bm{\alpha}_k \D \bm{N}_t \right) \bm{N}_t^{\top} + \bm{\alpha}_k \mathrm{Dg} (\E [\bm{\lambda}_t]) \D t
	\end{align*}
	and
	\begin{align*}
		\frac{\D \E [\bm{\lambda}_{k,t} \bm{N}_t^{\top}]}{\D t} &= \E[\bm{\lambda}_{k,t} \bm{\lambda}^{\top}_t]  - \bm{\beta}_{k} \E[\bm{\lambda}_{k,t} \bm{N}_t^{\top}] + \bm{\alpha}_k \E[ \bm{\lambda}_t \bm{N}_t^{\top}]  + \bm{\alpha}_k \mathrm{Dg} (\E [\bm{\lambda}_t])
	\end{align*}
	and
	\begin{align*}
		\A_k = \E[\bm{\lambda}_{k}(t) \bm{\lambda}^{\top}_t] - \bm{\beta}_{k}(\A_k t + \B_k) + \bm{\alpha}_k (\A t + \B ) + \bm{\alpha}_k \mathrm{Dg} (\E [\bm{\lambda}_t])
	\end{align*}
	and
	\begin{align*}
		\A_k &= \bm{\beta}_k^{-1}\bm{\alpha}_k \A \\
		\B_k &= \bm{\beta}_k^{-1}(- \A_k + \bm{\alpha}_k \B + \E[\bm{\lambda}_{k}(t)\bm{\lambda}_{t}^{\top}] + \bm{\alpha}_k \mathrm{Dg} (\E [\bm{\lambda}_t] ) ) \\
		&=\bm{\beta}_k^{-1}(- \bm{\beta}_k^{-1}\bm{\alpha}_k \A  + \bm{\alpha}_k \B + \E[\bm{\lambda}_{k}(t)\bm{\lambda}_{t}^{\top}] + \bm{\alpha}_k \mathrm{Dg} (\E [\bm{\lambda}_t] ) ).
	\end{align*}
	Substituting the above to Eq.~\eqref{Eq:ABk}, we obtain the desired result.
\end{proof}

\begin{proof}[Proof for Propositon~\ref{Prop:second}]
	Using Lemma~\ref{Lemma:quad} with
	$ \bm{a} = \bm{b} = 0, \bm{f}_x = \bm{f}_y = \bm{\mathrm{I}}$
	and Proposition~\ref{Prop:lN}, we obtain the result.
\end{proof}

\section{Estimation results}

\begin{table}[!hbt]
	\small
	\caption{Estimates and standard errors under the two-kernel model for January 2018 based on the NBBO of IBM}
	\centering
	\begin{tabular}{cccccccc}
		\hline
		date     &   $\mu$  &  $\alpha_{1s}$ & $\alpha_{1c}$ & $\alpha_{2s}$ & $\alpha_{2c}$  & $\beta_1$ & $\beta_2$ \\
		\hline
		2018-01-02 & 0.1073 & 683.0 & 229.2 & 3.429 & 4.353 & 2041 & 46.34 \\
		& (0.0017) & (0.0596)  & (0.1365) & (0.0347) & (0.0771) & (0.0500) & (0.1181) \\
		2018-01-03 & 0.1697 & 637.0 & 215.3  & 4.773 & 6.473 & 1707 & 56.46 \\
		& (0.0021) & (0.0111) & (0.0392) & (0.0261) & (0.0464) & (0.0049)  & (0.0352) \\
		2018-01-04 & 0.1295 & 797.5 & 260.7 & 7.451 & 9.918 & 2423 & 81.54 \\
		& (0.0018) & (0.0850) & (0.0762) & (0.0421) & (0.0964) & (0.0386) & (0.1340) \\
		2018-01-05 & 0.1193  & 804.3 & 200.2 & 4.314 & 5.815 & 2434 & 46.56 \\
		& (0.0018) & (0.1918) & (0.1414) & (0.2975) & (0.1434) & (0.2182) & (0.0872) \\
		2018-01-08 & 0.1223 & 806.1 & 218.1 & 4.633  & 6.917 & 2430 & 51.03 \\
		& (0.0018) & (0.0605) & (0.0879) & (0.0394) & (0.0560) & (0.0153) & (0.0693) \\
		2018-01-09 & 0.1129 & 707.7 & 179.5 & 3.519 & 3.869 & 2182   & 38.15  \\
		& (0.0017) & (0.0851) & (0.0237)  & (0.1310) & (0.3077) & (0.0891) & (0.0997) \\
		2018-01-10 & 0.0747 & 575.1 & 186.1 & 3.315 & 4.344 & 2006 & 40.49 \\
		& (0.0014) & (0.3245) & (0.3192) & (0.1709) & (0.2196) & (0.0404) & (0.1087) \\
		2018-01-11 & 0.0918  & 737.8 & 200.5 & 3.537 & 5.074  & 2094 & 45.51 \\
		& (0.0015) & (0.0893) & (0.0980) & (0.0330) & (0.1472) & (0.0844) & (0.0306) \\
		2018-01-12 & 0.1033 & 608.9 & 176.5 & 5.976 & 7.822 & 1755 & 64.47 \\
		& (0.0016) & (0.0746) & (0.1005) & (0.0481) & (0.2354) & (0.0312) & (0.0941) \\
		2018-01-16 & 0.1778   & 484.1  & 186.6 & 4.066 & 6.259   & 1582 & 46.20 \\
		& (0.0021) & (0.0530) & (0.0377) & (0.0327) & (0.0576) & (0.0181) & (0.0488) \\
		2018-01-17 & 0.1815 & 552.5 & 192.1 & 5.098 & 8.305 & 1777  & 60.78 \\
		& (0.0023) & (0.0381) & (0.1950) & (0.1334) & (0.1306) & (0.0661) & (0.1500) \\
		2018-01-18 & 0.2074 & 660.1 & 207.4 & 5.144 & 5.520 &  1922 & 54.13 \\
		& (0.0023) & (0.0288) & (0.0242) & (0.0554) & (0.0470) & (0.0137) & (0.0209) \\
		2018-01-19 & 0.2851 & 602.5 & 310.7 & 2.597 & 3.306 & 7.392 & 36.94 \\
		& (0.0027) & (0.2681) & (0.2784) & (0.2357) & (0.0352) & (0.2185 ) & (0.0155) \\
		2018-01-22 & 0.1443 & 662.8 & 290.4 & 2.333 & 4.713 & 1791 & 44.51 \\
		& (0.0019) & (0.0343) & (0.0358) & (0.0424) & (0.0164) & (0.0210) & (0.0648) \\
		2018-01-23 & 0.1250 & 474.6 & 172.4 & 2.835 & 5.142 & 1657 & 38.61 \\
		& (0.0018) & (0.1237) &	(0.1455) & (0.2019) & (0.1499) & (0.0533) & (0.1033) \\
		2018-01-24 & 0.2082 & 576.3 & 178.0 & 2.506 & 6.525 & 2183 & 33.79 \\
		& (0.0024) & (0.0528) & (0.0775) & (0.0588) & (0.0717) & (0.0404) & (0.1154) \\
		2018-01-25 & 0.1656 & 485.5 & 116.1 & 2.658 & 8.710 & 2130 & 36.02 \\
		& (0.0021) & (0.0863) & (0.0879) & (0.0960) & (0.0717) & (0.0537) & (0.0372) \\
		2018-01-26 & 0.0873 & 585.7 & 140.9 & 4.259 & 7.836 & 2093 & 54.11 \\
		& (0.0015) & (0.1618) & (0.1657) & (0.0982) & (0.2668) & (0.1149) & (0.1861) \\
		2018-01-29 & 0.1543 & 522.4 & 105.8 & 1.974 & 8.045 & 2209 & 31.83 \\
		& (0.0021) & (0.0347) & (0.1528) & (0.1201) & (0.1694) & (0.2561) & (0.0945) \\
		2018-01-30 & 0.1676 & 566.0 & 138.7 & 2.614 & 7.365 & 2085 & 35.06\\
		& (0.0021) & (0.0369) & (0.0625) & (0.0331) & (0.0574) & (0.0223) & (0.0281) \\
		2018-01-31 & 0.1274 & 651.3 & 141.3 & 3.326 & 6.150 & 2073 & 36.84\\
		& (0.0018) & (0.0635) & (0.0241) & (0.0723) & (0.0443) & (0.0900) & (0.0579)\\
		\hline
	\end{tabular}
\end{table}

\afterpage{%
	\clearpage
	\thispagestyle{empty}
	\begin{landscape}
		\centering 
					\footnotesize
		\captionof{table}{Maximum likelihood estimates by the Hawkes models with different number of kernels based on NBBO of IBM for first four days of 2018}\label{Table:4kernel}
		\begin{tabular}{cccccccccccccccc}

			\hline
			date & kernel & $\mu$  &  $\alpha_{1s}$ & $\alpha_{1c}$ & $\alpha_{2s}$ & $\alpha_{2c}$ & $\alpha_{3s}$ & $\alpha_{3c}$ &  $\alpha_{4s}$ & $\alpha_{4c}$  & $\beta_1$ & $\beta_2$ & $\beta_3$ & $\beta_4$ & llh \\
			\hline
			01-02 & one & 0.1306 & 323.8 & 130.4 & & & & & & & 854.9 & & & & 9967 \\
			& & (0.0018) & (0.0537) & (0.0777) & & & & & & & (0.0151) & & & &\\
			& two & 0.1073 & 683.1 & 229.2 & 3.429 & 4.354 & & & & & 2041 & 46.34 & & & 12748  \\
			& & (0.0017) & (0.0596) & (0.1365) & (0.0347) & (0.0771) & & & & & (0.0500) & (0.1181) & & & \\
			& three & 0.0757 & 639.1 & 206.5 & 4.861 & 6.501 & 0.0632 & 0.0324 & & & 1947 & 74.10 & 0.7097 & &  13210\\
			& & (0.0018) & (0.1463) & (0.1246) & (0.1470) & (0.0827) & (0.0041) & (0.0040) & & & (0.0496) & (0.0863) & (0.0395) & &  \\
			& four & 0.0884 &	609.5 & 220.2 & 7.515 & 8.118 & 0.2603 & 0.6074 & 0.0794 & 0.0425 & 1982 & 121.9 & 17.76 & 1.405 & 13105 \\
			& & (0.0028) & (0.1943) & (0.2937) & (0.1724) & (0.0567) & (0.0350) & (0.1174) & (0.0193) & (0.0127) & (0.1912) & (0.1162) & (0.0191) & \\
			\hline
			01-03 & one & 0.2127 & 301.2 & 127.0 & & & & & & & 699.3 & & & & 35526 \\
			& & (0.0026) & (0.1107) & (0.1127) & & & & & & & (0.0329) & & & & \\
			& two & 0.1694 & 637.0 & 215.3 & 4.773 & 6.473 & & & & & 1707 & 56.46 & & & 41300 \\
			& & (0.0021) & (0.0111) & (0.0392) & (0.0261) & (0.0464) & & & & & (0.0049) & (0.0352) & \\
			& three & 0.1217 & 804.4 & 252.4 & 8.694 & 10.99 & 0.0982 & 0.0407 & & & 2322 & 93.87 & 1.221 & & 42155 \\
			& & (0.0023) & (0.0309) & (0.0185 ) & (0.0282) & (0.0054) & (0.0045) & (0.0036) & & & (0.0197) & (0.0426) & (0.0038) \\
			& four & 0.1279 & 1189 & 400.5 & 240.3 & 85.94 &  5.2708 & 8.0202 &0.1150 & 0.0502 & 8220 & 1113 & 77.42 & 1.664 &  42500 \\	
			& & (0.0022) & (0.2202) & (0.1362) & (0.1831) & (0.0589) & (0.0684) & (0.1180) &  (0.0056) & (0.0053) & (0.0405) & (0.1558) & (0.1408) &  (0.0173) \\
			\hline
			01-04 & one & 0.1539 & 298.6 & 138.9 & & & & & & & 749.4 & & & & 18304 \\
			& & (0.0020) & (0.1674) & (0.1570) & & & & & & & (0.1553) \\
			& two & 0.1295 & 797.5 & 260.7 & 7.451 & 9.918 & & & & & 2423 & 81.54 & & & 21984 \\
			& & (0.0018) & (0.0851) & (0.0762) & (0.0421) & (0.0964) & & & & & (0.3862) & (0.1340)\\
			& three & 0.0882 & 722.7 & 235.5 & 8.443 & 12.17 & 0.1361 & 0.0239 & & &  2203 & 106.7 & 1.226 & & 22706 \\
			& & (0.0021) & (0.1882) & (0.1390) & (0.0154) & (0.0192) & (0.0063) & (0.0046) & & & (0.1251) & (0.0707) & (0.0740) \\
			& four & 0.0909 & 1311 & 386.0 & 245.6 & 99.23 & 0.1477 & 0.0233 & 4.664 & 8.283 & 9772 & 1125 & 1.423 & 82.49 & 23060 \\
			& & (0.0018) & (0.0522) & (0.1534) & (0.0258) & (0.0754) & (0.0062) & (0.0046) & (0.0477) & (0.0622) & (0.0377) & (0.0455) & (0.0336) & (0.0214) \\
			\hline
			01-05 & one & 0.1516 & 291.2 & 104.9 & & & & & & & 747.0 & & & & 10118 \\
			& & (0.0024) & (0.6086) & (1.9505)  & & & & & & & (0.3064)   \\
			& two & 0.1193 & 804.3 & 200.2 & 4.314 & 5.815 & & & & & 2434 & 46.56 & & & 14405 \\
			& & (0.0018) & (0.1918) & (0.1414) & (0.2975) & (0.1434) & & & & & (0.2182) & (0.0872) \\
			& three & 0.0827 & 831.1 & 199.4 & 5.459 & 7.532 & 0.1012 & 0.0353  & & & 2551 & 64.05 & 0.9769 & & 15000 \\
			& & (0.0023) & (0.1007) & (0.2396) & (0.0500) & (0.1007) & (0.0072) & (0.0048) & & &  (0.0606) & (0.0920) & (0.0850) \\
			& four & 0.0839 & 1023 & 176.2 & 131.1 & 51.38 & 3.087 & 5.564 & 0.1115 & 0.0393 &  4934 & 926.9 &  51.37 & 1.151  & 15129 \\
			& & (0.0019) & (0.0808)	& (0.0970) & (0.0991) & (0.1765) & (0.0813) & (0.0990) & (0.0057) & (0.0047) &  (0.1392) & (0.0954) & (0.0802) & (0.0475) & (0.0117) \\
			\hline
		\end{tabular}
		
	\end{landscape}
	\clearpage
}

\end{document}